\documentclass[preprint2]{aastex63}

\usepackage{acronym}
\acrodef{DESI}{Dark Energy Spectroscopic Instrument}
\acrodef{FoV}{field-of-view}
\acrodef{FWHM}{full width at manuscripthalf maximum}
\acrodef{GAMA}{Galaxy And Mass Assembly}
\acrodef{IMF}{initial mass function}\acrodef{M/L}{mass-to-light ratio}
\acrodef{J-PAS}{Javalambre-Physics of the Accelerated Universe Astrophysical Survey}
\acrodef{J-PLUS}{Javalambre Photometric Local Universe Survey}
\acrodef{LSB}{low surface brightness}
\acrodef{MAP}{maximum a posteriori probability}
\acrodef{NUTS}{No U-Turn Sampler}
\acrodef{SNR}{signal-to-noise ratio}
\acrodef{SED}{spectral energy distributions}
\acrodef{SDSS}{Sloan digital sky survey}
\acrodef{SMUDGes}{Systematically Measuring Ultra-diffuse Galaxies}
\acrodef{S-PLUS}{Southern Photometric Local Universe Survey}
\acrodef{SSP}{single stellar population}
\acrodef{UDGs}{Ultra-diffuse galaxies}
\received{January 7, 2020}
\revised{February 7, 2020}
\accepted{February 12, 2020}
\submitjournal{ApJS}

\shorttitle{100 SMUDGes in S-PLUS}
\shortauthors{Barbosa et al.}
\graphicspath{{./}{figures/}}
\begin{document}

\title[100 SMUDGes in S-PLUS]{One hundred SMUDGes in S-PLUS: ultra-diffuse galaxies flourish in the field}

\correspondingauthor{C. E. Barbosa}
\email{kadu.barbosa@gmail.com}

\author[0000-0002-5292-2782]{C. E. Barbosa} 
\affiliation{Steward Observatory, University of Arizona, 933 N Cherry Ave, Tucson, AZ 85719, USA}
\affiliation{Universidade de S\~{a}o Paulo, Instituto de Astronomia, Geof\'isica e Ci\^encias Atmosf\'ericas, Departamento de Astronomia, Rua do Mat\~{a}o 1225, S\~{a}o Paulo, SP, 05508-090, Brazil}

\author[0000-0002-5177-727X]{D. Zaritsky}
\affiliation{Steward Observatory, University of Arizona, 933 N Cherry Ave, Tucson, AZ 85719, USA}

\author[0000-0001-7618-8212]{R. Donnerstein}
\affiliation{Steward Observatory, University of Arizona, 933 N Cherry Ave, Tucson, AZ 85719, USA}

\author[0000-0002-0123-9246]{H. Zhang}
\affiliation{Steward Observatory, University of Arizona, 933 N Cherry Ave, Tucson, AZ 85719, USA}

\author[0000-0002-4928-4003]{A. Dey}
\affiliation{NSF's National Optical-Infrared Astronomy Research Laboratory, P.O. Box 26732, Tucson, AZ 85726, USA}

\author[0000-0002-5267-9065]{C. Mendes de Oliveira}
\affiliation{Universidade de S\~{a}o Paulo, Instituto de Astronomia, Geof\'isica e Ci\^encias Atmosf\'ericas, Departamento de Astronomia, Rua do Mat\~{a}o 1225, S\~{a}o Paulo, SP, 05508-090, Brazil}

\author{L. Sampedro}
\affiliation{Universidade de S\~{a}o Paulo, Instituto de Astronomia, Geof\'isica e Ci\^encias Atmosf\'ericas, Departamento de Astronomia, Rua do Mat\~{a}o 1225, S\~{a}o Paulo, SP, 05508-090, Brazil}

\author{A. Molino}
\affiliation{Universidade de S\~{a}o Paulo, Instituto de Astronomia, Geof\'isica e Ci\^encias Atmosf\'ericas, Departamento de Astronomia, Rua do Mat\~{a}o 1225, S\~{a}o Paulo, SP, 05508-090, Brazil}

\author{M. V. Costa-Duarte}
\affiliation{Universidade de S\~{a}o Paulo, Instituto de Astronomia, Geof\'isica e Ci\^encias Atmosf\'ericas, Departamento de Astronomia, Rua do Mat\~{a}o 1225, S\~{a}o Paulo, SP, 05508-090, Brazil}

\author[0000-0003-1846-4826]{P. Coelho}
\affiliation{Universidade de S\~{a}o Paulo, Instituto de Astronomia, Geof\'isica e Ci\^encias Atmosf\'ericas, Departamento de Astronomia, Rua do Mat\~{a}o 1225, S\~{a}o Paulo, SP, 05508-090, Brazil}

\author[0000-0002-0620-136X]{A. Cortesi}
\affiliation{Universidade de S\~{a}o Paulo, Instituto de Astronomia, Geof\'isica e Ci\^encias Atmosf\'ericas, Departamento de Astronomia, Rua do Mat\~{a}o 1225, S\~{a}o Paulo, SP, 05508-090, Brazil}

\author[0000-0001-7907-7884]{F. R. Herpich}
\affiliation{Universidade de S\~{a}o Paulo, Instituto de Astronomia, Geof\'isica e Ci\^encias Atmosf\'ericas, Departamento de Astronomia, Rua do Mat\~{a}o 1225, S\~{a}o Paulo, SP, 05508-090, Brazil}

\author{J. A. Hernandez-Jimenez}
\affiliation{Universidade de S\~{a}o Paulo, Instituto de Astronomia, Geof\'isica e Ci\^encias Atmosf\'ericas, Departamento de Astronomia, Rua do Mat\~{a}o 1225, S\~{a}o Paulo, SP, 05508-090, Brazil}
\affiliation{Universidad Andr\'es Bello, Departamento de Ciencias F\'isicas, Fern\'andez Concha 700, Las Condes, Santiago, Chile}

\author{T. Santos-Silva}
\affiliation{Universidade de S\~{a}o Paulo, Instituto de Astronomia, Geof\'isica e Ci\^encias Atmosf\'ericas, Departamento de Astronomia, Rua do Mat\~{a}o 1225, S\~{a}o Paulo, SP, 05508-090, Brazil}

\author[0000-0002-1564-2933]{E. Pereira}
\affiliation{Universidade de S\~{a}o Paulo, Instituto de Astronomia, Geof\'isica e Ci\^encias Atmosf\'ericas, Departamento de Astronomia, Rua do Mat\~{a}o 1225, S\~{a}o Paulo, SP, 05508-090, Brazil}

\author[0000-0002-4382-8081]{A. Werle}
\affiliation{Universidade de S\~{a}o Paulo, Instituto de Astronomia, Geof\'isica e Ci\^encias Atmosf\'ericas, Departamento de Astronomia, Rua do Mat\~{a}o 1225, S\~{a}o Paulo, SP, 05508-090, Brazil}
\affiliation{Universidade Federal de Santa Catarina, Departamento de F\'isica, SC 88040-900, Brazil, Florian\'opolis, SC 88040-900, Brazil}

\author[0000-0002-8214-7617]{R. A. Overzier}
\affiliation{Universidade de S\~{a}o Paulo, Instituto de Astronomia, Geof\'isica e Ci\^encias Atmosf\'ericas, Departamento de Astronomia, Rua do Mat\~{a}o 1225, S\~{a}o Paulo, SP, 05508-090, Brazil}
\affiliation{Observat\'orio Nacional, Minist\'erio da Ci\^encia, Tecnologia, Inova\c c\~ao e Comunica\c c\~oes, Rua General Jos\'e Cristino, 77, S\~ao Crist\'ov\~ao, 20921-400 Rio de Janeiro, RJ, Brazil}

\author[0000-0001-9672-0296]{R. Cid Fernandes}	
\affiliation{Universidade Federal de Santa Catarina, Departamento de F\'isica, SC 88040-900, Brazil, Florian\'opolis, SC 88040-900, Brazil}

\author{A. V. Smith Castelli}
\affiliation{Instituto de Astrof\'isica de La Plata, UNLP, CONICET, Paseo del Bosque s/n, B1900FWA La Plata, Argentina}
\affiliation{Facultad de Ciencias Astron\'omicas y Geof\'isicas, UNLP, Paseo del Bosque s/n, B1900FWA, La Plata, Argentina}

\author{T. Ribeiro}
\affiliation{NSF's National Optical-Infrared Astronomy Research Laboratory, P.O. Box 26732, Tucson, AZ 85726, USA}
\affiliation{Universidade Federal de Sergipe, Departamento de F\'isica, Av. Marechal Rondon, S/N, 49000-000 S\~ao Crist\'ov\~ao, SE, Brazil}

\author[0000-0002-4064-7234]{W. Schoenell}
\affiliation{GMTO Corporation, 465 N. Halstead Street, Suite 250, Pasadena, CA 91107, USA}
\affiliation{Universidade Federal do Rio Grande do Sul (UFRGS), Instituto de F\'isica, Departamento de Astronomia, Av. Bento Gon\c calves 9500, Porto Alegre, RS, Brazil}

\author{A. Kanaan}
\affiliation{Universidade Federal de Santa Catarina, Departamento de F\'isica, SC 88040-900, Brazil, Florian\'opolis, SC 88040-900, Brazil}



\begin{abstract} 
We present the first systematic study of the stellar populations of ultra-diffuse galaxies (UDGs) in the field, integrating the large area search and characterization of UDGs by the SMUDGes survey with the twelve-band optical photometry of the S-PLUS survey. Based on Bayesian modeling of the optical colors of UDGs, we determine the ages, metallicities and stellar masses of 100 UDGs distributed in an area of $\sim 330$ deg$^2$ in the Stripe 82 region. We find that the stellar masses and metallicities of field UDGs are similar to those observed in clusters and follow the trends previously defined in studies of dwarf and giant galaxies. However, field UDGs have younger luminosity-weighted ages than do UDGs in clusters. We interpret this result to mean that field UDGs have more extended star formation histories, including some that continue to form stars at low levels to the present time. Finally, we examine stellar population scaling relations that show that UDGs are, as a population, similar to other low-surface brightness galaxies.
\end{abstract}


\section{Introduction} \label{sec:intro}

%The
\ac{UDGs} are a recently defined class of \ac{LSB} galaxy initially found in large numbers in the Coma Cluster \citep{2015ApJ...798L..45V}. Their unusually large half-light radii, $R_e\ge 1.5$ kpc, for galaxies with such low central surface brightness, $\mu_{0,g}\ge 24$ mag arcsec$^{-2}$, are striking. Although large LSB galaxies have been known for quite some time \citep{1976Natur.263..573D, 1984AJ.....89..919S, 1988ApJ...330..634I, 1988AJ.....95.1389S, 1995MNRAS.275..121S, 1997AJ....114..635D, 1997ApJ...482..104S}, the current excitement originates from indications, either from kinematic measures of the unresolved light or globular clusters in these galaxies \citep{2016ApJ...819L..20B, 2018ApJ...856L..31T, 2019ApJ...880...91V}, or from the numbers of globular clusters alone \citep{2016ApJ...822L..31P, 2016ApJ...830...23B, 2017ApJ...844L..11V, 2018ApJ...856L..30V}, that at least some \ac{UDGs} lie in massive $(>10^{11}$ M$_\odot$) halos.

The detection of populations of \ac{UDGs} in galaxy clusters \citep[e.g.][]{2015ApJ...798L..45V, 2015ApJ...809L..21M, 2016A&A...590A..20V, 2017A&A...608A.142V, 2017ApJ...846...26S} led to the exploration of a possible evolutionary link between these galaxies and their harsh environment \citep{2017ApJ...850...99S, 2018RNAAS...2...43C, 2018MNRAS.480L.106O, 2018ApJ...866L..11B, 2019MNRAS.485..382C}. However, \ac{UDGs} were also found in less dense environments, such as filaments \citep[e.g..][]{2016AJ....151...96M}, groups \citep[e.g.][]{2015A&A...581A..82M, 2016A&A...596A..23S, 2017MNRAS.468.4039R, 2017A&A...607A..79V, 2018Natur.555..629V}, the field \citep[e.g.][]{2017ApJ...842..133L, 2018ApJ...866..112G} and even voids \citep{2019MNRAS.486..823R}. Moreover, observational studies \citep[e.g.][]{2015MNRAS.452..937Y, 2017MNRAS.464L.110Z, 2018MNRAS.473.3747S, 2018MNRAS.475.4235A} and theoretical ones \citep[e.g.][]{2016MNRAS.459L..51A, 2017MNRAS.466L...1D, 2017MNRAS.470.4231R, 2018MNRAS.478..906C, 2019MNRAS.tmp.2566L, 2019MNRAS.487.5272J} found that \ac{UDGs} span a wide range of physical properties, and perhaps a correspondingly large range of origin stories.

A key challenge in developing a unified understanding of \ac{UDGs} and their relation to other galaxies is that the data so far come from disparate studies, with different selection criteria, and mostly focus on high density environments. These deficits are exacerbated by the difficulties posed in observing such low surface brightness galaxies. Photometric information, such as
broadband colors \citep[e.g.][]{2019MNRAS.488.2143P}, are available for many \ac{UDGs} but are of limited value in determining the properties of the stellar populations, while spectroscopy, which can provide the necessary information, is only available for a small number of galaxies \citep[e.g.][]{2016AJ....151...96M, 2017ApJ...838L..21K, 2018MNRAS.478.2034R, 2018MNRAS.479.4891F, 2018ApJ...859...37G}. 

Recently, \citet{2019ApJS..240....1Z} presented the initial results from the \ac{SMUDGes} survey, a systematic study to detect and characterize the photometric properties of \ac{UDGs} over a large area of the sky ($\sim$14000 deg$^2$) using data from the Legacy survey \citep{2019AJ....157..168D}. In its initial release, \ac{SMUDGes} provided a catalog containing 275 UDG candidates, including most of the galaxies previously reported within 10$^{\circ}$ of the Coma cluster by \citet{2015ApJ...798L..45V} and \citet{2016ApJS..225...11Y}, using a relatively small area of the total survey ($334$ deg$^2$). SMUDGes has now analyzed the SDSS Stripe 82 region and identified 172 candidate \ac{UDGs} in this region, which is a much more typical region of the sky than that around the Coma cluster (Zaritsky et al. in prep.).

The limited passbands of the Legacy survey preclude stellar population modeling and spectroscopic observations of the SMUDGes candidates will always be highly limited \citep[Kadowaki et al., in prep]{2017ApJ...838L..21K}. Interestingly, the requirement of large \ac{FoV}, multi-passband imaging in the study of \ac{UDGs} intersects with the interest of several ongoing cosmological surveys, such as the \acl{J-PAS} \citep[\acs{J-PAS},][]{2014arXiv1403.5237B}\acused{J-PAS}, the \acl{J-PLUS} \citep[J-PLUS,][]{2019A&A...622A.176C}\acused{J-PLUS} and the \acl{S-PLUS} \citep[\acs{S-PLUS},][]{2019MNRAS.489..241M}\acused{S-PLUS}. Here we explore the synergy between \ac{SMUDGes} ands \ac{S-PLUS} to perform the first statistical study of the stellar populations of UDG candidates over an area of sky that is not dominated by high density environments. Despite limited overlap within Stripe 82 between the two surveys, we were able to study a sample of 100 UDG candidates and perform the largest detailed population study of these galaxies to date. 

This paper is structured as follows. In \S\ref{sec:data}, we describe the two datasets used in this work. In \S\ref{sec:photometry}, we describe the photometry of the UDG candidates in the context of S-PLUS, and in \S\ref{sec:sedfitting}, we present the method developed to determine their stellar populations using a Bayesian framework. In \S\ref{sec:discussion}, we present our results and discuss the main implications of our work for our understanding of the nature of UDGs. We conclude and summarize this work in \S\ref{sec:conclusion}. Throughout, we assume a standard $\Lambda$CDM cosmology whenever necessary, assuming $H_0=70$ km s$^{-1}$ Mpc$^{-1}$. All magnitudes use the AB system \citep{1964ApJ...140..689O, 1983ApJ...266..713O}.

\section{Data}
\label{sec:data}

\subsection{SMUDGes sample}

Our ability to locate \ac{LSB} galaxies have been limited both by the lack of sensibility and instrumental constraints, and various attempts have been made to optimize observations at low surface brightness \citep[e.g.][]{2001ApJS..137..117G, 2015ApJ...809L..21M, 2014PASP..126...55A}. However, there have been no systematic attempt to use current, large volume archival data to search for \ac{LSB} galaxies, which have not been identified before because standard pipelines are not optimized to find such systems. The \ac{SMUDGes} project \citep{2019ApJS..240....1Z} was conceived to develop an automatized way to search for \ac{LSB} galaxies over a large area of the sky using data from the Legacy imaging survey \citep{2019AJ....157..168D}, a deep three-band observational campaign that supports the \ac{DESI} project \citep{2011arXiv1106.1706S, 2016arXiv161100036D, 2016arXiv161100037D}.

Most of what is known about \ac{UDGs} as a population is based on observations of the Coma cluster \citep{2015ApJ...798L..45V, 2016ApJS..225...11Y}. In this initial stage of the project, the \ac{SMUDGes} detection algorithm has been constrained to search \ac{UDGs} similar to those found in Coma, and thus is focused on systems with angular sizes $R_e \gtrsim 5$ arcsec, which are easier to classify than smaller objects in the absence of redshift information. The methodology used to identify \ac{UDGs} is described in detail in \citet{2019ApJS..240....1Z} and Zaritsky et al. (in prep.), and here we summarize the main steps of the process. First, bright, saturated sources are detected, modeled and replaced in the \ac{DESI} images by background noise, whereas fainter background and foreground sources are carefully modeled and subtracted. Then, wavelet filtering is used to detect sources according to size and surface brightness criteria, defined to have $\mu_{0,g} \ge 24$ mag arcsec$^{-2}$ and $R_e > 5.3$\arcsec. Finally, all UDG candidates are modeled with a single S\'ersic component using \textsc{GALFIT} \citep{2002AJ....124..266P, 2010AJ....139.2097P}. 

In this work, we use a sample of 172  UDG candidates in the Stripe 82 area. From SMUDGes we adopt the values of $m_g, m_r, m_z,$ S\'ersic index $n$, and $R_e$ in arcsec, position angle and axis ratio. Without distance estimates, we cannot determine whether these systems pass the common defining criteria for UDGs, $R_e \ge 1.5$ kpc, and some of these galaxies may actually be dwarf galaxies at small distances. The redshift by association (defining high density regions in terms of normal galaxies and assigning SMUDGes to the redshift of that overdensity) worked for 25 candidates in Stripe 82 and all 25 satisfy the $R_e > 1.5$ kpc criterion at the assigned distance. Only 1 has $R_e > 6$ kpc, which seems to be about the upper limit on size - it has $R_e = 8.6$ kpc - which suggests that this one may have the wrong redshift. In this particular case, the UDG candidate is close in projection to a nearby bright galaxy and so it may instead be a satellite of that galaxy (Zaritsky et al., in prep). Therefore, we work under the hypothesis that we have a sample of UDGs with low contamination by dwarf galaxies, but we examine this issue again further below.

\subsection{S-PLUS DR1 data}

We use data from the \ac{S-PLUS} first data release (DR1), which covers an area of 336 deg$^{\text{2}}$ in the Stripe 82 equatorial field, observed with the T80S, a 0.8m robotic telescope with a wide \ac{FoV} of $\sim 1.8$ deg$^{\text{2}}$, located in Cerro Tololo, Chile. The DR1 data are already reduced and are publicly available in the NSF’s National Optical-Infrared Astronomy Research Laboratory archive\footnote{\url{https://datalab.noao.edu/splus/index.php}}. Details about the survey strategy and data reduction process are described by \citet{2019MNRAS.489..241M} while the photometric calibration is described in Sampedro et al. (in prep.).

The main survey strategy is aimed at obtaining %a 
large coverage of the Southern sky ($\sim 9000$ deg$^{\text{2}}$) for astronomical and cosmological studies in the local universe. The S-PLUS uses the same photometric system of the J-PLUS survey \citep{2012SPIE.8450E..3SM}, which consists of twelve optical bands, including 5 broad-bands similar to those used by the \ac{SDSS} $ugriz$ system, and a set of seven narrow-band ($\Delta\lambda= 100-200$\r{A}) filters placed at various rest-frame optical features, including [OII] ($\lambda_{\text{eff}}=3771$ \r{A}), Ca H+K ($\lambda_{\text{eff}}=3941$ \r{A}), H$\delta$ ($\lambda_{\text{eff}}=4094$ \r{A}), G-band ($\lambda_{\text{eff}}=4292$ \r{A}), Mg $b$ triplet ($\lambda_{\text{eff}}=5133$ \r{A}), H$\alpha$ ($\lambda_{\text{eff}}=6614$ \r{A}) and the Ca triplet ($\lambda_{\text{eff}}=8611$ \r{A}). Considering a \ac{SNR} threshold of 3, the survey is complete in the broad bands to magnitudes of $u=21.07$, $g=21.79$, $r=21.6$, $i=21.22$ and $z=20.64$, whereas it is complete to magnitudes of $\sim20.4$ in all narrow bands \citep{2019MNRAS.489..241M}.

\begin{figure*}[ht!]
\epsscale{1.15} % 2 columns
\plotone{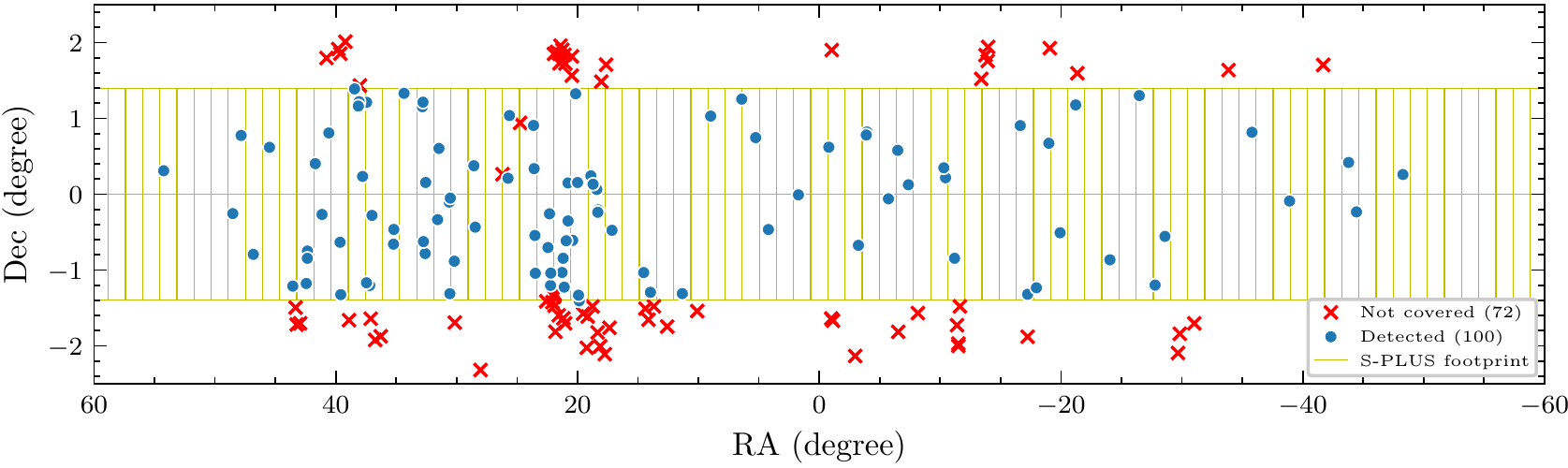}
\caption{Distribution of the SMUDGes Stripe 82 UDG candidates on the sky and their overlap with S-PLUS. 
The grid lines represent the S-PLUS footprint for the Stripe 82 area, red crosses indicate all SMUDGes galaxies that are outside the footprint or fell within gaps between exposures, and filled circles indicate the location of all UDG candidates with S-PLUS data.
\label{fig:footprint}}
\end{figure*}

The S-PLUS DR1 data cover Stripe 82 using a pair of exposures at each right ascension, limiting the declination to the range $-1.4^\circ\leq\text{dec}\leq+1.4^\circ$. Moreover, the SPLUS DR1 data did not use dithering, causing occasional gaps between exposures, resulting in a few UDGs that are not observed despite being within the footprint of the survey. In total, we have observations for only 100 SMUDGes from the initial sample of 172. Figure \ref{fig:footprint} shows the spatial distribution of the SMUDGes Stripe 82 sample overlapped with the S-PLUS DR1 footprint. 

\section{Photometry of UDGs from the S-PLUS data}
\label{sec:photometry}

\ac{UDGs} are not easily detected given the surface brightness limits of the S-PLUS survey, and only one UDG was previously detected in the DR1 catalog of photometric redshifts \citep{2019arXiv190706315M}. Therefore, we had to obtain our own photometry of the \ac{UDGs} from the S-PLUS images leveraging the information from the deeper SMUDGes photometry.

Regarding the data quality of S-PLUS, all images in the S-PLUS Main Survey, which includes Stripe 82, are obtained during photometric nights with seeing $\le$ 2\arcsec. Among the 61 different tiles used in this work, the mean \ac{FWHM} over all bands is 1.4\arcsec. Moreover, because each field is imaged in all bands consecutively in a given observational block, there are only small seeing variations among all bands for each tile (mean standard deviation among bands of 0.14\arcsec). We conclude that there is no need to homogenize the seeing across the images for our photometry.  

For each UDG, we perform aperture photometry in each of the 12 bands from S-PLUS using the \textsc{photutils} package \citep{Bradley_2019_2533376}. To ensure consistent photometry, for each UDG we define an elliptical aperture with a semi-major axis length of $R_e$, and location, position angle and ellipticity determined from the \textsc{GALFIT} S\'ersic profile fitting from the SMUDGes analysis. We subtract local sky using an elliptical annulus with inner and outer radii of $2.5R_e$ and $4R_e$, respectively. Presuming that the S\'ersic profile is a good approximation to the surface brightness profile of the UDGs, this annulus  is large enough to avoid contamination of the sky region by the galaxy itself \citep[see][]{2005PASA...22..118G}. We use sigma-clipping to remove the contribution of other sources when we estimate the median background. All observed magnitudes are corrected for the foreground Galactic extinction using the dust maps from \citet{1998ApJ...500..525S} recalibrated by \citet{2011ApJ...737..103S} assuming that $R_V=3.1$ for the Milky Way \citep{1979ARA&A..17...73S}.

The aperture photometry method described above has the advantage of allowing the detection of most \ac{UDGs} in individual bands despite their low \ac{SNR}. However, in most cases (87 galaxies), at least one band was not detected, as the measured flux inside the galaxy is smaller than the flux in the sky annulus. In these cases, we are only able to set an upper limit on the source flux. There are missing detections in most of the bands, but the blue bands are the most affected, in particular the narrow bands F378 and F395, for which there are flux detections in only $\sim 60$\% of the galaxies. Nevertheless, in the majority of the cases (97 galaxies), we have flux detections in at least 6 bands, which already provides better spectral coverage in the optical than do the SDSS bands, and 80\% of the galaxies have detected flux in at least 9 bands.

\begin{figure*}
\gridline{
\fig{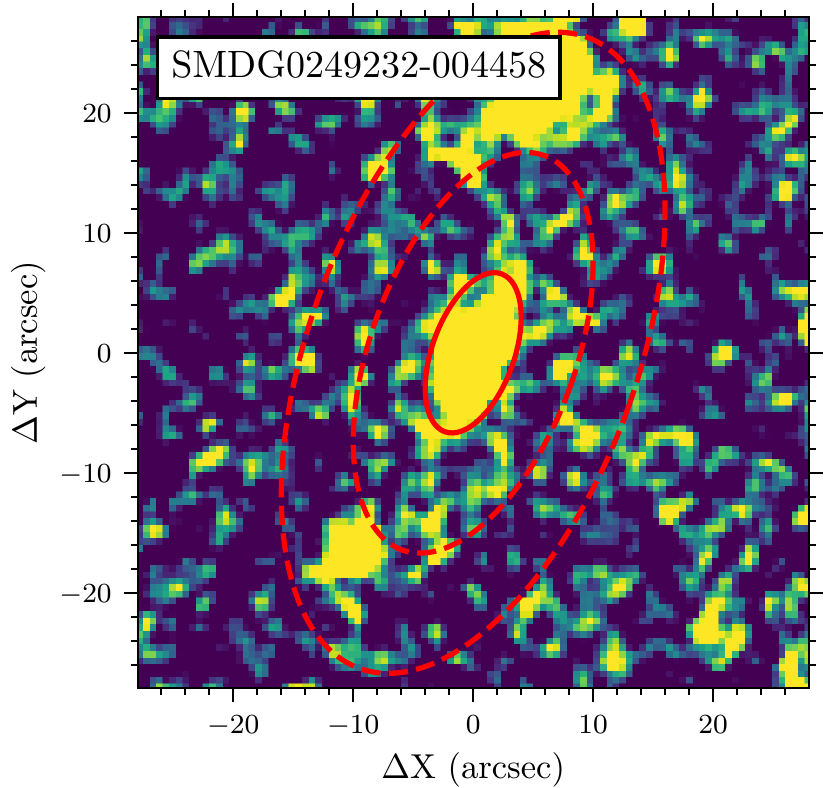}{0.33\textwidth}{}
\fig{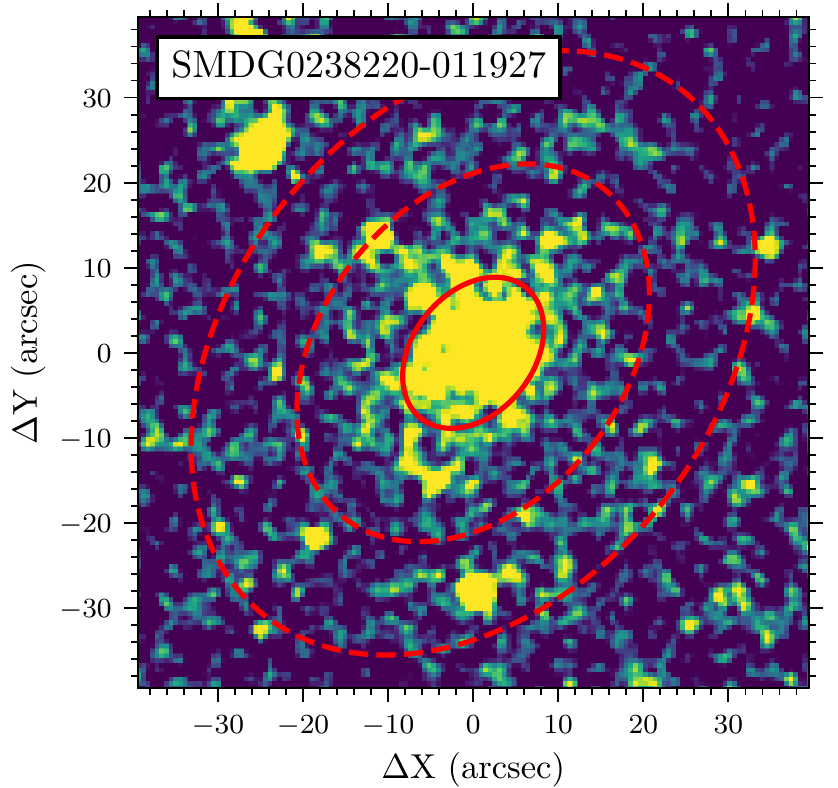}{0.33\textwidth}{}
\fig{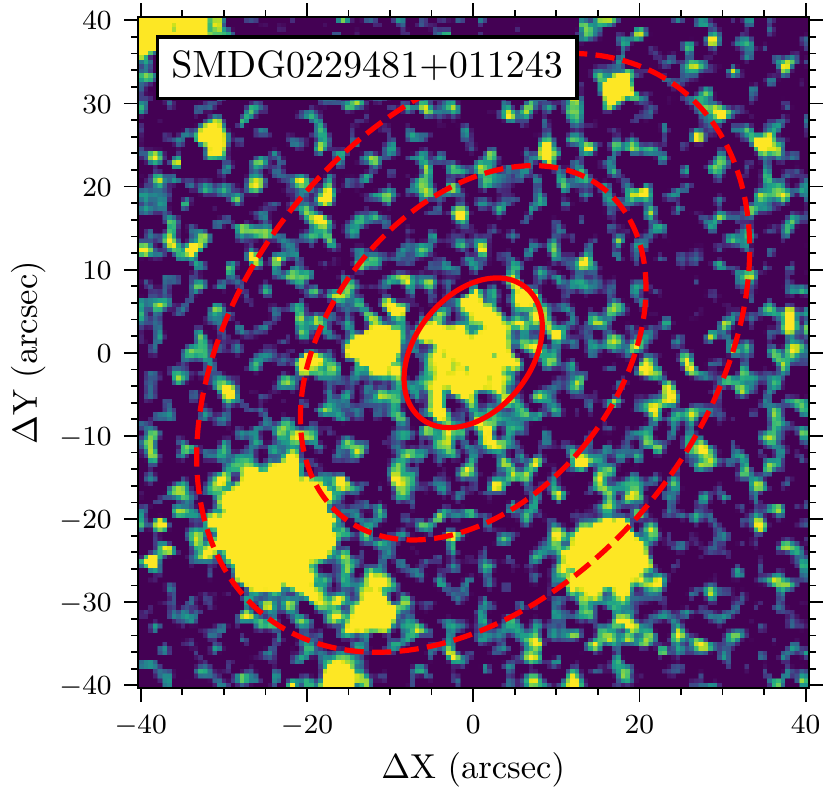}{0.33\textwidth}{}
          }
\gridline{
\fig{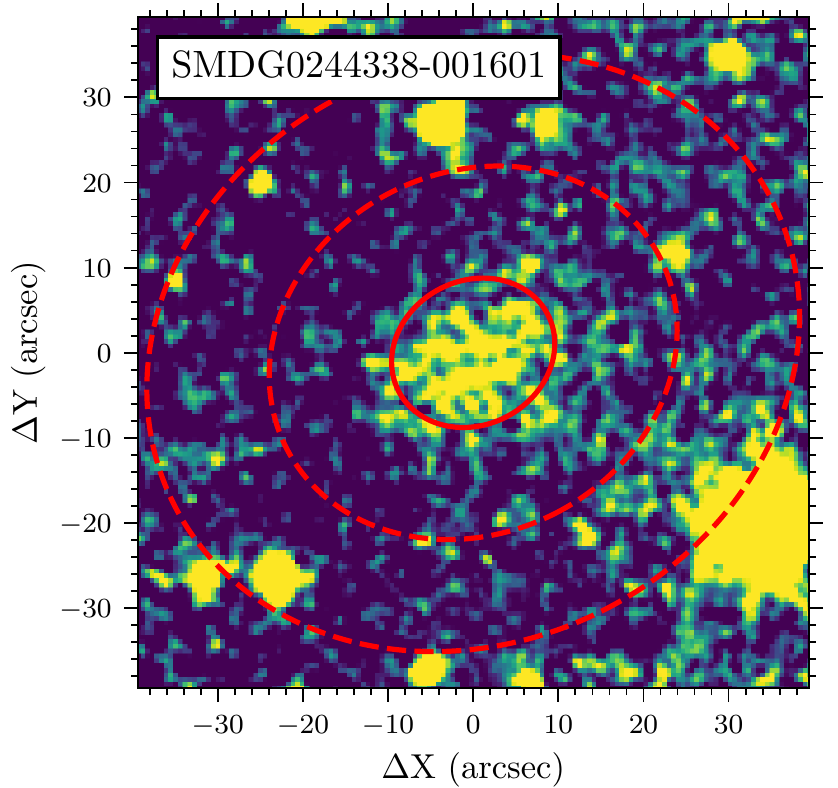}{0.33\textwidth}{}
\fig{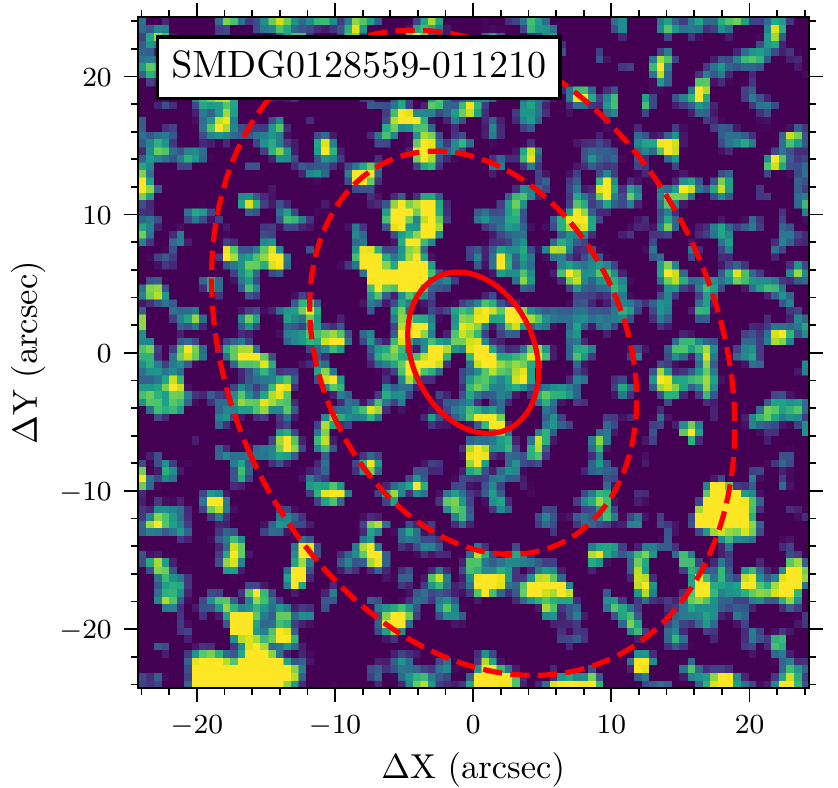}{0.33\textwidth}{}
\fig{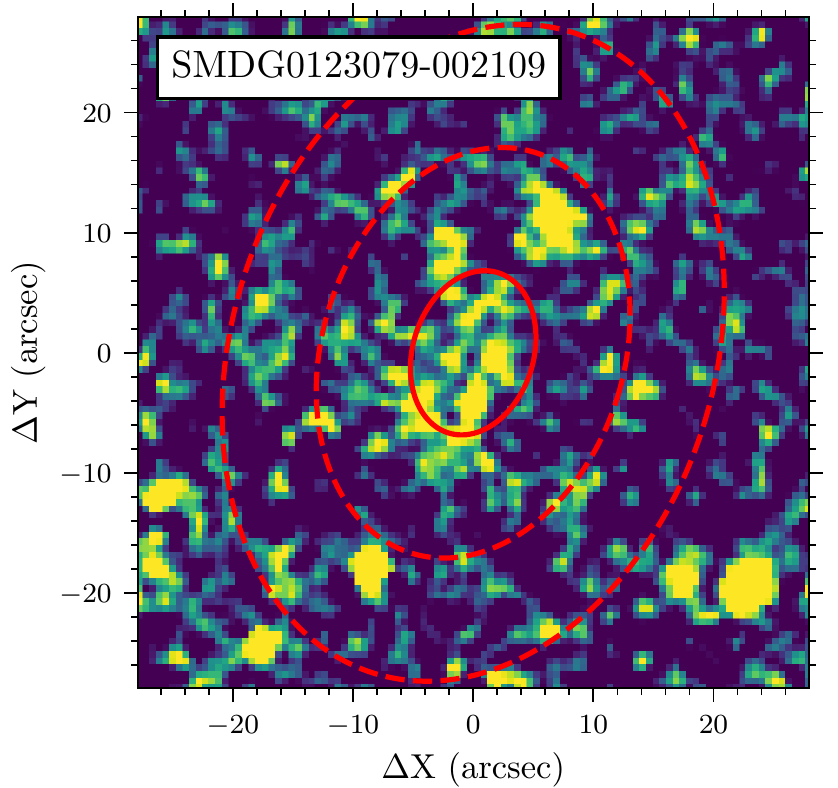}{0.33\textwidth}{}
          }
\caption{Six example detection images of SMUDGes galaxies in S-PLUS. The panels are ordered according to decreasing $g$-band central surface brightness, starting in the upper left. The solid red ellipses outline the $1R_e$ photometric aperture while the dashed red lines indicate the annulus used for background estimation. \label{fig:detection_images}}
\end{figure*}

In Figure \ref{fig:detection_images}, we show a sample of detection images of SMUDGes UDGs, produced by stacking all 12 S-PLUS bands, ordered in decreasing central surface brightness in the $g$ band, $\mu_{0,g}$, and highlight the photometric apertures.

\section{Stellar populations from multi-band observations}
\label{sec:sedfitting}

We quantify the properties of the stellar populations of our sample by performing \ac{SED} fitting of the galaxies in all detected bands of the \ac{S-PLUS} data. Considering that the star formation history (SFH) of galaxies is difficult to determine from photometric data alone, and that simulations indicate that UDGs may have bursty SFHs \citep{2017MNRAS.466L...1D, 2018MNRAS.478..906C}, we assume that \ac{SED}s may be described by a \ac{SSP}, such that 

\begin{equation}
f_\lambda(\lambda) = f_0 \cdot \text{SSP}(\text{[Fe/H]}, \text{Age}, z)10^{{-0.4 A_\lambda}}
\end{equation}

\noindent where $f_0$ is a scale factor for the spectral flux density of the galaxy, $\text{SSP}$ represents a single stellar population model that depends of the metallicity ([Fe/H]) and age, and the redshift $z$ of the galaxy, and $A_\lambda$ represents a dust-screen attenuation model. One important cautionary point about the use of \acp{SSP} to represent a potentially more complicated SFH is that the derived properties are luminosity weighted. As appreciated previously \citep[cf.][]{2007MNRAS.374..769S}, luminosity-weighted ages are expected to be biased toward the youngest populations, in contrast to the luminosity-weighted metallicity, which reflects more closely the mass-weighted average.   

Considering both the low \ac{SNR} of the observations and the low spectral resolution of the photometric system, we expect that derived parameters may be correlated, as is the case in the well-known age-metallicity degeneracy problem \citep{1994ApJS...95..107W}, and that some parameters will not be properly estimated. Therefore, we use a Bayesian statistical model to fit the \ac{SED} of the galaxies and to estimate the stellar population parameters. Using this approach, we can identify possible parameter correlations and  provide uncertainties that are marginalized over the distribution of all the other parameters in the model. 

Bayes' theorem allows for the inference of the probability distribution of a set of parameters $\theta$ in a statistical model based on a dataset $D$ using the relation 

\begin{equation}
 p(\theta | D) \propto p(\theta) p(D|\theta)\text{,}
\end{equation}

\noindent where $p(\theta|D)$ represents the posterior probability distribution of the parameters $\theta$ given the data $D$, $p(\theta)$ represents the prior distribution of the parameters, and $p(D|\theta)$ is the likelihood distribution \citep[see, e.g.][]{gelmanbda04}. Below we describe the priors for all of the parameters in our model.

\subsection{Prior and likelihood distributions}

The flux scale factor $f_0$ can vary by orders of magnitude depending on the brightness of the source. Therefore, it is more convenient to model its logarithm, which can be described by the prior

\begin{equation}
 \log f_0 \sim \textrm{Normal}(\mu_{0}, \sigma_{0}^2)\text{,}
\end{equation}

\noindent where $\mu_{0}$ and $\sigma_{0}^2$ indicate the mean and the variance of the distribution respectively. In practice, we estimate $\mu_{0}$ using the data, and we assume $\sigma_{0}=3$ to allow a large range of magnitudes. 

Our modeling is parameterized in terms of two stellar population parameters, the age and metallicity, whose priors are set by the limits of the model ranges. In this work, we use the E-MILES models \citep{2016MNRAS.463.3409V}, assuming prior distributions given by

\begin{eqnarray}
 \text{[Fe/H]}(\text{dex}) \sim \textrm{Uniform}(-1.79, 0.4)\\  
 \text{Age} (\text{Gyr}) \sim \textrm{Uniform}(0.4, 14)\text{.}
\end{eqnarray}

\noindent The main reason to set the limits above is to ensure that the \ac{SSP} models have good quality in the ultra-violet according to the classification of \citet{2010MNRAS.404.1639V}, resulting in the exclusion of SSP models with [Fe/H]$=-2.27$, which may not be appropriate for metallicity estimation. Additionally, we also require a regular grid in the parameter space to perform linear interpolation of the SSP models, allowing a continuous coverage of ages and metallicities. As a consequence, we had to restrict the models to ages greater than 0.4 Gyr because part of the young SSP models are not extended to the near-infrared, which is necessary to cover the \ac{S-PLUS} $z$ band properly. In particular, we adopt \ac{SSP} models with bimodal \ac{IMF}, a piecewise function defined by \citet{1996ApJS..106..307V} that matches the Salpeter \ac{IMF} for masses $>0.8M_\odot$ but is flattened at lower masses similarly to the Milky Way \ac{IMF} \citep[e.g.][]{2003PASP..115..763C}. 
Given that the current version of the E-MILES stellar population models do not include the abundance of individual or alpha elements yet, we are restricted to the base models, which assume that [M/H]=[Fe/H] at solar metallicity. However, this assumption does not hold at low metallicities because the Milky Way stars used in the computation of the models are themselves alpha-enhanced \citep[see][]{2010MNRAS.404.1639V}. The consequences of possible offsets resulting from non-solar abundance ratios are discussed further below.

The  redshifts  of  our galaxies are  of  great  interest because they set the distances to the galaxies and their physical parameters, and allow a proper classification of the candidates as UDGs. Without additional spectroscopic or redshift-by-association for our sample, we consider a prior that takes into consideration a few assumptions. \ac{LSB} galaxies with angular sizes $R_e\gtrsim 5$\arcsec{} have only been associated to environments with distances smaller than 100 Mpc \citep[see][]{2018A&A...620A.166G}, thus we can assume all UDG candidates are nearby. Moreover, all UDG candidates were selected with a minimum effective radius of $R_e=5.3$\arcsec, which implies a physical radius of $R_e=2.5$ kpc at the distance of Coma, $100$ Mpc, or a redshift of $z\approx 0.023$. At a distance as low as $200$ Mpc, or $z\approx 0.046$, these UDGs would all already have an effective radius of $R_e \ge 5$ kpc, which is larger than most UDGs found so far \citep[e.g.][]{2017A&A...608A.142V}. We conclude that it is very unlikely that many of our candidates lie at $z > 0.04$. Based on these considerations, we use the prior

\begin{equation}
 z\sim \textrm{HalfNormal}(0.03^2)\text{,}
\end{equation}

\noindent where we adopt the half-normal distribution to restrict the redshifts to positive values, and we assume a variance of $0.03^2$. In practice, this prior implies a median redshift $z\approx 0.02$, similar to Coma, with peak probability at $z=0$. 

Regarding the dust attenuation, our data include only wavelengths $\lambda > 3000$ \r{A} for low redshift galaxies, avoiding the 2175 \r{A} bump \citep{1965ApJ...142.1683S}. For these wavelengths, most of the extinction laws, such as those obtained for the Milky Way \citep{1976asqu.book.....A, 1986ApJ...307..286F}, the Large Magellanic Cloud \citep{1986ApJ...307..286F}, the Small Magellanic Cloud \citep{1984A&A...132..389P, 1985A&A...149..330B}, and starburst galaxies \citep{2000ApJ...533..682C}, agree \citep[see][]{2019MNRAS.483.2382W}. We adopt a parametrized extinction law from \citet{1989ApJ...345..245C}, which depends on two parameters, the total extinction in the $V$-band, $A_V$, and the total-to-selective extinction, $R_V$. The total extinction is modeled according to the prior

\begin{equation}
 A_V \sim \textrm{Exponential}(0.2)\text{,}
\end{equation}

\noindent where $0.2$ represents the mean value of the exponential distribution. This prior restricts the value of the extinction to positive values and also favors smaller extinction values rather than large. We also allow $R_V$ to vary in our models using the prior

\begin{equation}
 R_V \sim \textrm{Normal}^+(3.1, 1.)\mbox{,}
\end{equation}

\noindent which assumes that the total-to-selective extinction is similar to that measured generally within the Milky Way \citep{1979ARA&A..17...73S}. The plus signal indicates that we restrict $R_V$ to positive values.

Finally, it is necessary to define a log-likelihood for the use of the Bayes' theorem. The widely common assumption is that the observed \ac{SED} consists of independent, normal random variables, and thus the log-likelihood can be simply described as a $\chi^2$-distribution. However, the accuracy of the model determined using the normal assumption may be compromised if the observations contain outliers \citep[see][]{Vanhatalo2009}. In observational settings, the causes of outliers may be either  external to the source, such as contamination by cosmic rays or the incomplete removal of sky, or internal to the source, as is the case when the model is incomplete, for example when it does not account for emission lines. 

Emission lines have been directly observed in optical observations of at least one 
cluster UDG \citep{2017ApJ...838L..21K}, and may be common in field UDGs \citep{2017ApJ...842..133L}. Observationally, emission lines systematically inflate the observed fluxes in passbands in which they appear, an effect that is likely to be most noticeable in the bluer, narrow bands. However, the modeling of emission lines requires the inclusion of secondary stellar population with young ages ($<0.01$ Gyr) and/or post-asymptotic giant branch stars \citep{2017ApJ...840...44B} plus a prescription for nebular emission \citep[e.g.][]{1999astro.ph.12179F, 1999ApJS..123....3L}. To simplify the modeling, we instead adopt a robust fitting approach that may deal with outliers, including possibly emission lines, adopting a Student's $t$-distribution log-likelihood.

Similar to the normal distribution, the Student's $t$-distribution is a symmetric and bell-shaped distribution, but with long tails that allow for a non-negligible probability far from the center of the distribution \citep[see][]{gelmanbda04}. Assuming that we are modeling $N$ discrete bands in a given SED, the log-likelihood that we map is given by 

\begin{eqnarray}
 \ln p(D | \theta ) &=& 
 N\log \left [ \frac{\Gamma\left (\frac{\nu + 1}{2}\right )}{\sqrt{\pi (\nu-2)}\Gamma\left (\frac{\nu}{2} \right )}\right ] \nonumber \\ 
 &-& \frac{1}{2}\sum_{i=1}^{N}\log \sigma_{i}^2 \nonumber \\ 
 &-&\frac{\nu+1}{2}\sum_{i=1}^N \log \left [ 1 + \frac{\mu_{i}^2}{\sigma_{i}^2 (\nu-2)} \right ]\mbox{,}
 \label{eq:llf}
\end{eqnarray}

\noindent where $\Gamma(x)$ represents the gamma function of variable $x$, $\nu$ represents the degrees of freedom of the Student's $t$-distribution, $\sigma_{i}$ represents the uncertainties of a given \ac{SED} for the $i$-th band, and the mean $\mu_{i}$ represents the difference between the observed and the model SED. The Student's $t$-distribution log-likelihood does not depend solely on the data and its uncertainties, but also on the value of another variable, $\nu$, which controls the amount of weight on the tails of the Student's $t$-distribution. For instance, if $\nu\rightarrow2$, the tails of the distribution have more weight in the distribution, whereas if $\nu\rightarrow + \infty$, the distribution tends to a normal distribution. We also model the value of $\nu$ during the log-likelihood mapping assuming a non-informative prior for the degrees-of-freedom given by

\begin{equation}
 \nu\sim \textrm{Uniform}(2,50)\mbox{,}
\end{equation}

\noindent which is required to be open-ended only in the lower bounds to avoid the undefined likelihood that occurs if $\nu=2$. 

\subsection{Sampling and results}

To deploy the above SED fitting modeling in the context of the \ac{S-PLUS} project, we developed a Bayesian SED fitter (\textsc{BSF}, Barbosa, in prep.) as a general tool to model either SEDs or spectra of galaxies. \textsc{BSF} is written in the Python programming language based on the \textsc{pymc3} statistical package \citep{Salvatier2016}, which allows for the construction of general Bayesian models while abstracting the complex issues related to the actual modeling and sampling. An attractive feature of the  \textsc{pymc3} package, not found in other commonly adopted packages such as the \textsc{emcee} \citep{2013PASP..125..306F}, is the implementation of the \acl{NUTS} \citep[\acs{NUTS},][]{2011arXiv1111.4246H}, a Hamiltonian Monte Carlo sampler that has been shown to perform well in complex, multidimensional problems without the need of manual tuning. This sampler works better than other traditional samplers, such as the Metropolis-Hastings algorithm \citep{1970Bimka..57...97H} and the Gibbs sampler \citep{Gelfand1990}, in problems with highly-correlated variables in a space parameter with hundreds of dimensions \citep[see][]{2011arXiv1111.4246H}. 

\begin{figure*}
\gridline{
\fig{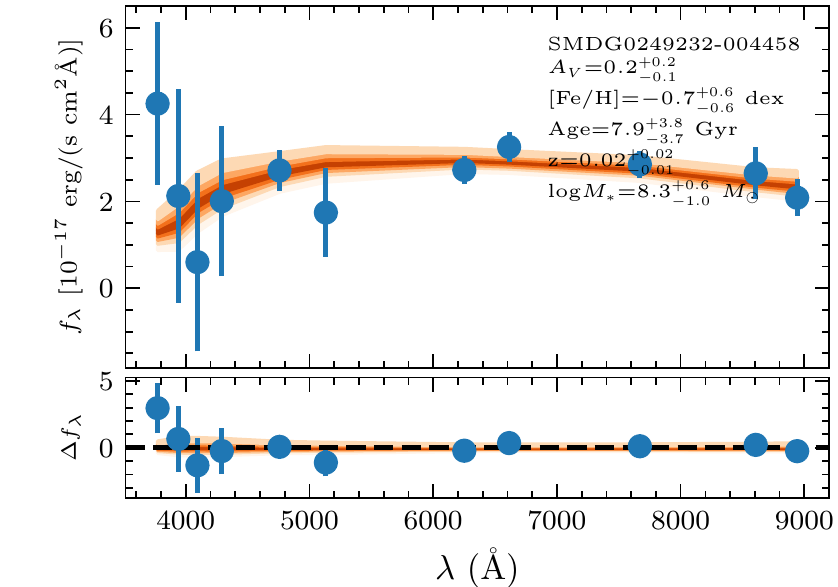}{0.45\textwidth}{}
\fig{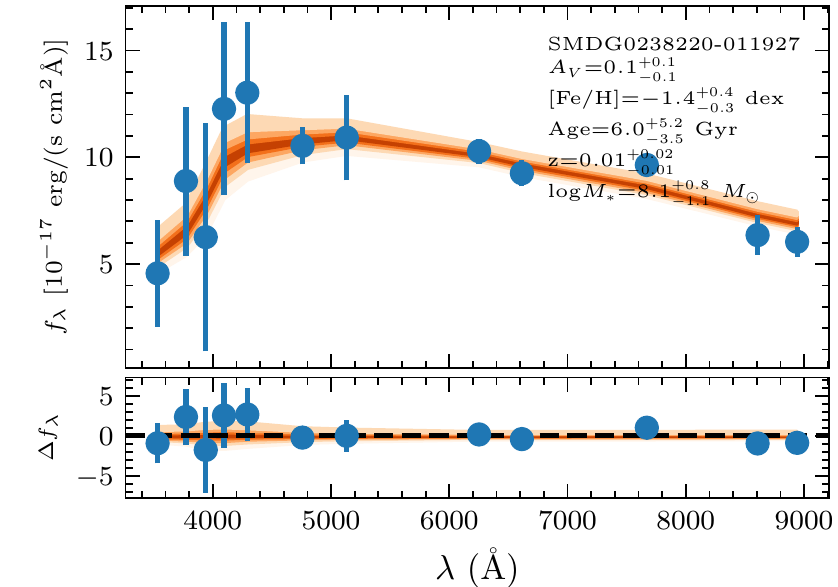}{0.45\textwidth}{}
          }
\gridline{
\fig{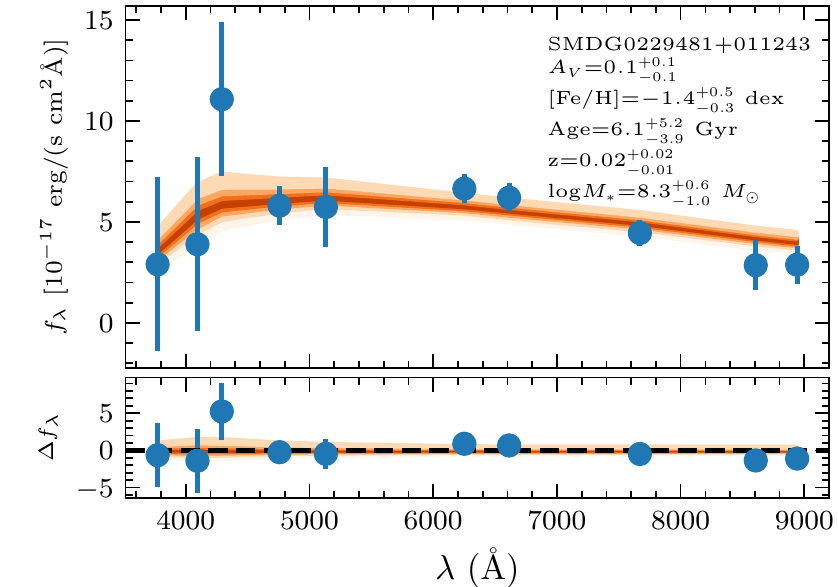}{0.45\textwidth}{}
\fig{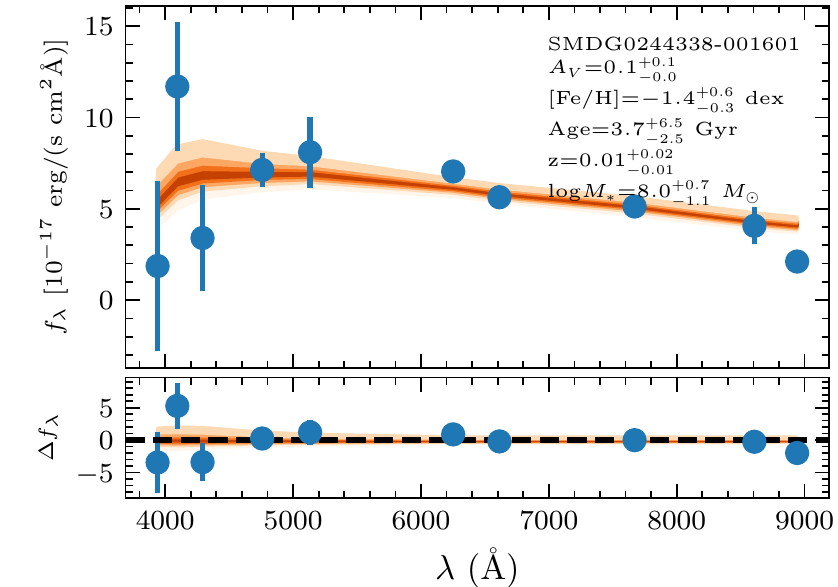}{0.45\textwidth}{}
          }
\gridline{
\fig{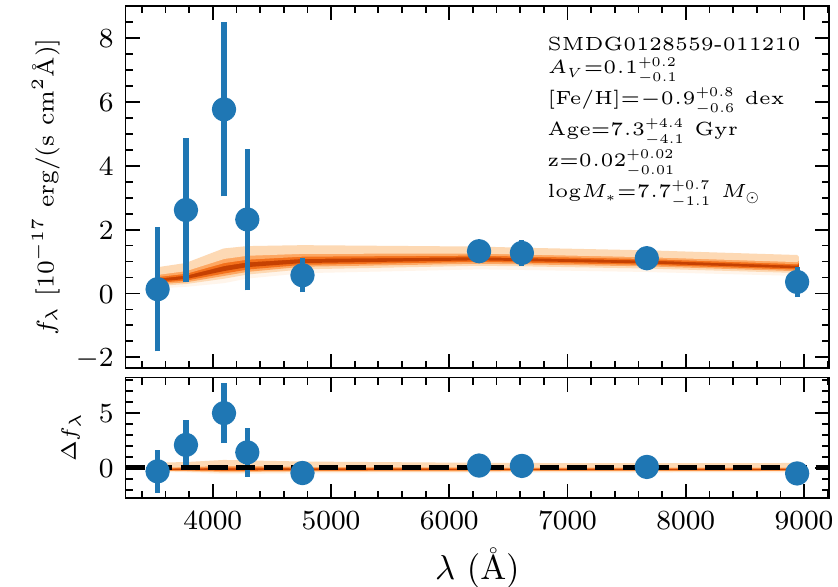}{0.45\textwidth}{}
\fig{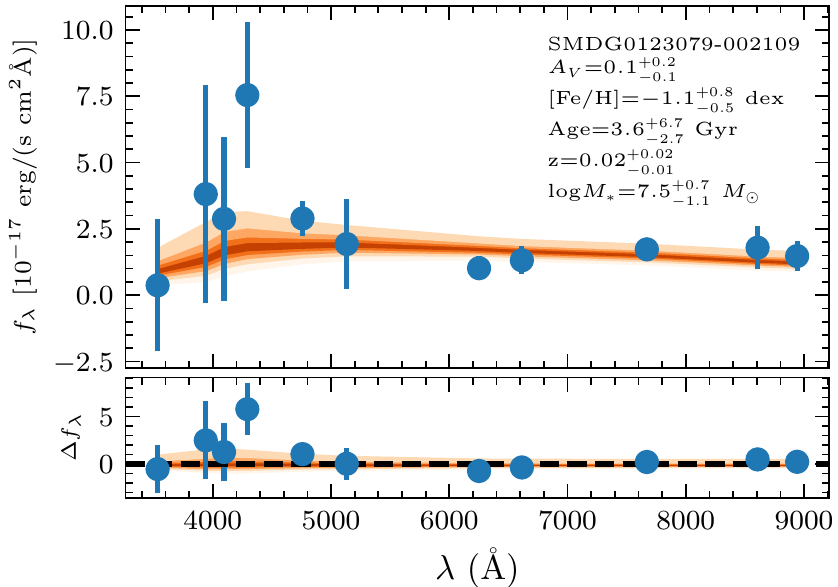}{0.45\textwidth}{}
          }
\caption{Resulting SED fits for the six examples presented in Fig.~\ref{fig:detection_images}. The blue filled circles represent the S-PLUS flux densities in detected bands. The shaded regions indicate the confidence percentile levels of the SED fitting, from 5 to 95\% in intervals of 10\%. Each panel comes in two portions, where the upper area shows the data and SED fit and the lower shows the residuals. A summary of the most relevant parameters is included in upper right of each panel. \label{fig:fitting_images}}
\end{figure*}

As discussed in \S\ref{sec:photometry}, there are a number of non-detections in our photometry. To simplify our modeling, we only included detected bands in the SED fitting for each UDG, and we leave the modeling including non-detections for forthcoming work. The samples from the posterior distributions were generated with \textsc{BSF} in four chains with 500 burn-in interactions and 500 draws. Figure \ref{fig:fitting_images} shows the comparison between the observations and the models for the sample of galaxies previously shown in Fig.~\ref{fig:detection_images}.

\begin{figure*}[ht]
\gridline{
\fig{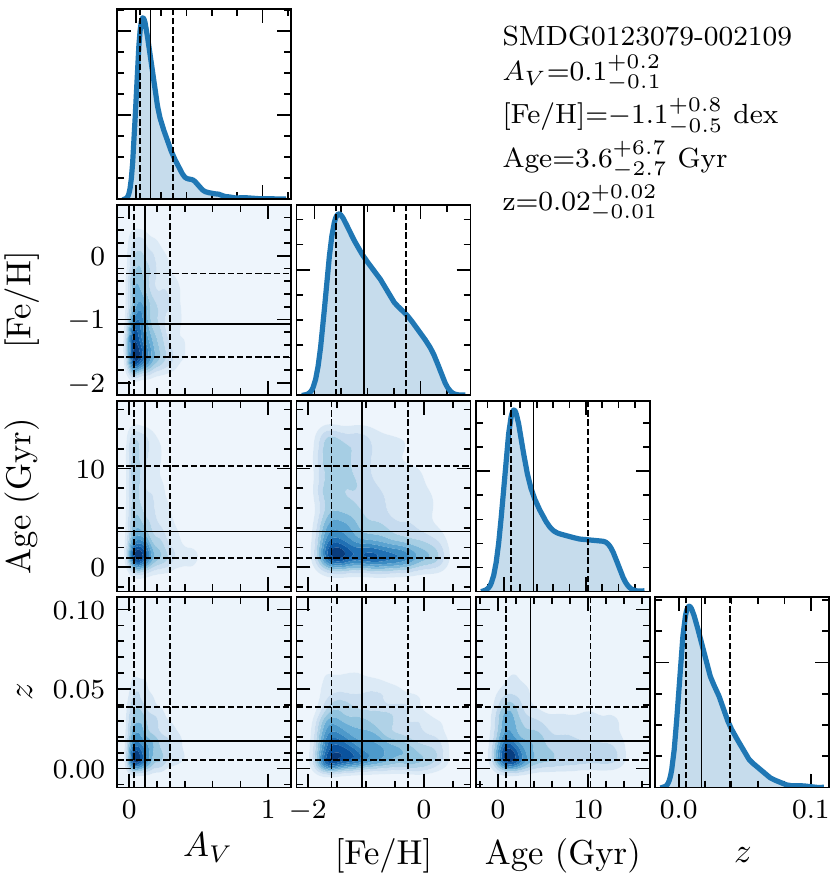}{0.45\textwidth}{}
\fig{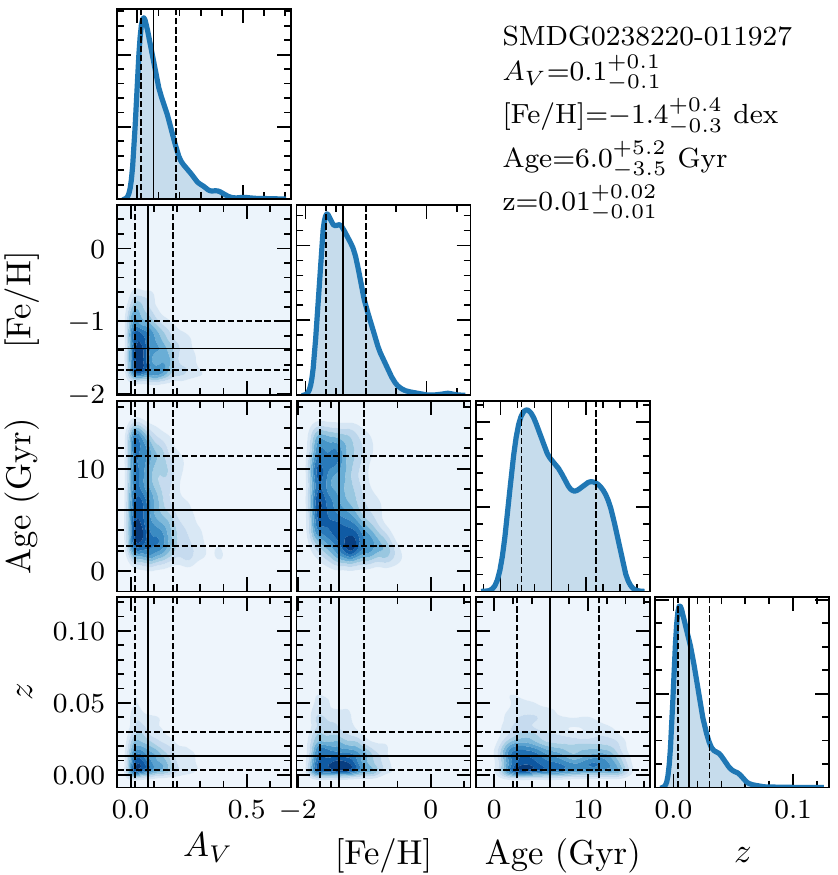}{0.45\textwidth}{}
          }
\caption{Sampled posterior distributions for two UDGs, SMDG0123079-002109 (left) and SMDG0238220-011927 (right) representing cases of the faint and bright end of the apparent magnitude distribution in our sample, respectively. The histograms along the diagonal contain the marginalized posterior distribution of parameters from where the resulting values and uncertainties are evaluated. The panels under the diagonal contain projections between pairs of variables, indicating how they are correlated. Solid lines mark the medians of the distributions, whereas dashed lines mark the 16\% and 84\% percentiles used to define the $1\sigma$ uncertainties. A summary of the results is included in the upper right corner of each panel. \label{fig:corner}}
\end{figure*}

To illustrate the process of obtaining representative values and uncertainties for the model parameters, Fig.~\ref{fig:corner} shows the posterior samples determined with \textsc{BSF} for two of the \ac{UDGs} presented in Fig.~\ref{fig:detection_images}: SMDG0123079-002109, representing a relatively faint galaxy and SMDG0238220-011927 representing a relatively bright galaxy. Throughout our analysis, we use the median to determine the representative value of all parameters, and we use the percentile values of 16\% and 84\% to estimate the $1\sigma$ confidence intervals of the parameters, always using the marginalized posterior distribution, shown in the histograms. In Table \ref{tab:results}, we present the results of this analysis for the first ten entries of the SMUDGes sample. The full table is available online in machine-readable format.

\begin{deluxetable*}{lccccc}
 \tablenum{1}
 \tablecaption{Stellar population parameters for SMUDGes UDGs obtained from SED fitting of the S-PLUS optical data. \label{tab:results}}
 \tablewidth{0pt}
 \tablehead{
\colhead{Name} & \colhead{$A_V$} & \colhead{[Fe/H] (dex)} & \colhead{Age (Gyr)}   & $z$ & \colhead{$\log M_\star$}}
\decimalcolnumbers
\startdata
SMDG0006543-000029 & $0.07_{-0.05}^{+0.12}$ & $-1.3_{-0.34}^{+0.7}$ & $2.8_{-2.1}^{+7.0}$ & $0.018_{-0.013}^{+0.020}$ & $7.5_{-1.1}^{+0.7}$ \\
SMDG0016502-002756 & $0.11_{-0.08}^{+0.17}$ & $-1.0_{-0.5}^{+0.8}$ & $7.1_{-4.0}^{+4.5}$ & $0.021_{-0.015}^{+0.020}$ & $8.0_{-1.0}^{+0.6}$ \\
SMDG0021031+004447 & $0.18_{-0.13}^{+0.28}$ & $-0.4_{-0.8}^{+0.6}$ & $8.0_{-4.5}^{+4.3}$ & $0.022_{-0.015}^{+0.021}$ & $7.9_{-0.9}^{+0.7}$ \\
SMDG0025396+011515 & $0.10_{-0.07}^{+0.19}$ & $-1.1_{-0.5}^{+0.8}$ & $5.8_{-4.5}^{+5.5}$ & $0.019_{-0.013}^{+0.020}$ & $7.5_{-1.0}^{+0.7}$ \\
SMDG0035569+010149 & $0.08_{-0.06}^{+0.14}$ & $-1.2_{-0.4}^{+0.6}$ & $3.1_{-2.3}^{+6.9}$ & $0.017_{-0.011}^{+0.021}$ & $7.4_{-1.0}^{+0.7}$ \\
SMDG0045200-011839 & $0.07_{-0.05}^{+0.12}$ & $-1.3_{-0.4}^{+1.0}$ & $1.5_{-0.9}^{+6.0}$ & $0.018_{-0.013}^{+0.018}$ & $7.4_{-1.2}^{+0.7}$ \\
SMDG0055526-011739 & $0.14_{-0.10}^{+0.24}$ & $-0.8_{-0.7}^{+0.8}$ & $7.4_{-4.6}^{+4.4}$ & $0.020_{-0.014}^{+0.022}$ & $7.5_{-1.1}^{+0.7}$ \\
SMDG0058071-010201 & $0.09_{-0.07}^{+0.14}$ & $-1.2_{-0.4}^{+0.8}$ & $4.6_{-3.5}^{+6.1}$ & $0.021_{-0.015}^{+0.021}$ & $8.0_{-1.0}^{+0.6}$ \\
SMDG0108359-002834 & $0.18_{-0.14}^{+0.28}$ & $-0.5_{-0.8}^{+0.6}$ & $8.0_{-4.4}^{+4.1}$ & $0.022_{-0.015}^{+0.024}$ & $7.9_{-1.1}^{+0.7}$ \\
SMDG0113101-001223 & $0.10_{-0.07}^{+0.15}$ & $-1.1_{-0.5}^{+0.8}$ & $4.3_{-3.4}^{+6.6}$ & $0.020_{-0.014}^{+0.020}$ & $7.4_{-1.1}^{+0.7}$ \\
\enddata
\tablecomments{Table sample containing only the first ten entries. The full table is available in machine-readable form.}
\end{deluxetable*}

\subsection{Stellar masses}

We determine the stellar mass of each UDG combining our SED fitting results of the S-PLUS data with the photometric properties measured in the deeper SMUDGes images. We adopt two different approaches. First, we use the SED fitting photometric redshift to estimate the distance and the total apparent $r$-band magnitude from SMUDGes to determine the total magnitude, assuming a Hubble-Lema\^itre law with $H_0=70\pm2$ km s$^{-1}$ Mpc$^{-1}$. Next, we use the $r$-band mass-to-light ratio from the E-MILES models \citep{2010MNRAS.404.1639V, 2012MNRAS.424..172R} to obtain the total stellar mass, assuming that $M_{\odot,r}=4.65$ \citep{2018ApJS..236...47W}. These calculations are performed using the Monte Carlo chains, and thus the uncertainties are marginalized over all parameters in the SED fitting model. Second, we estimate the stellar mass using the color-mass relation from  \citet{2011MNRAS.418.1587T}, which is based on data from the \ac{GAMA} survey \citep{2009A&G....50e..12D, 2011MNRAS.413..971D}. Reassuringly, the stellar masses resulting from the two approaches always agree to within $0.1$ dex, which is a difference that is much smaller than the typical mass uncertainties ($\sim$0.8 dex). We conclude that our stellar mass estimates are likely to be dominated by internal random uncertainties rather than by a systematic error in the conversion between luminosity and stellar mass. For the remainder of this work, we use the stellar masses determined using the first method. The stellar masses derived by the first method are also presented in Table \ref{tab:results}.

\subsection{Evaluating the posterior distributions}
\label{sec:posteriors}

\begin{figure}[ht!]
\epsscale{1.2} % 2 columns 
% \epsscale{0.6} % 1 column (manuscript)
\plotone{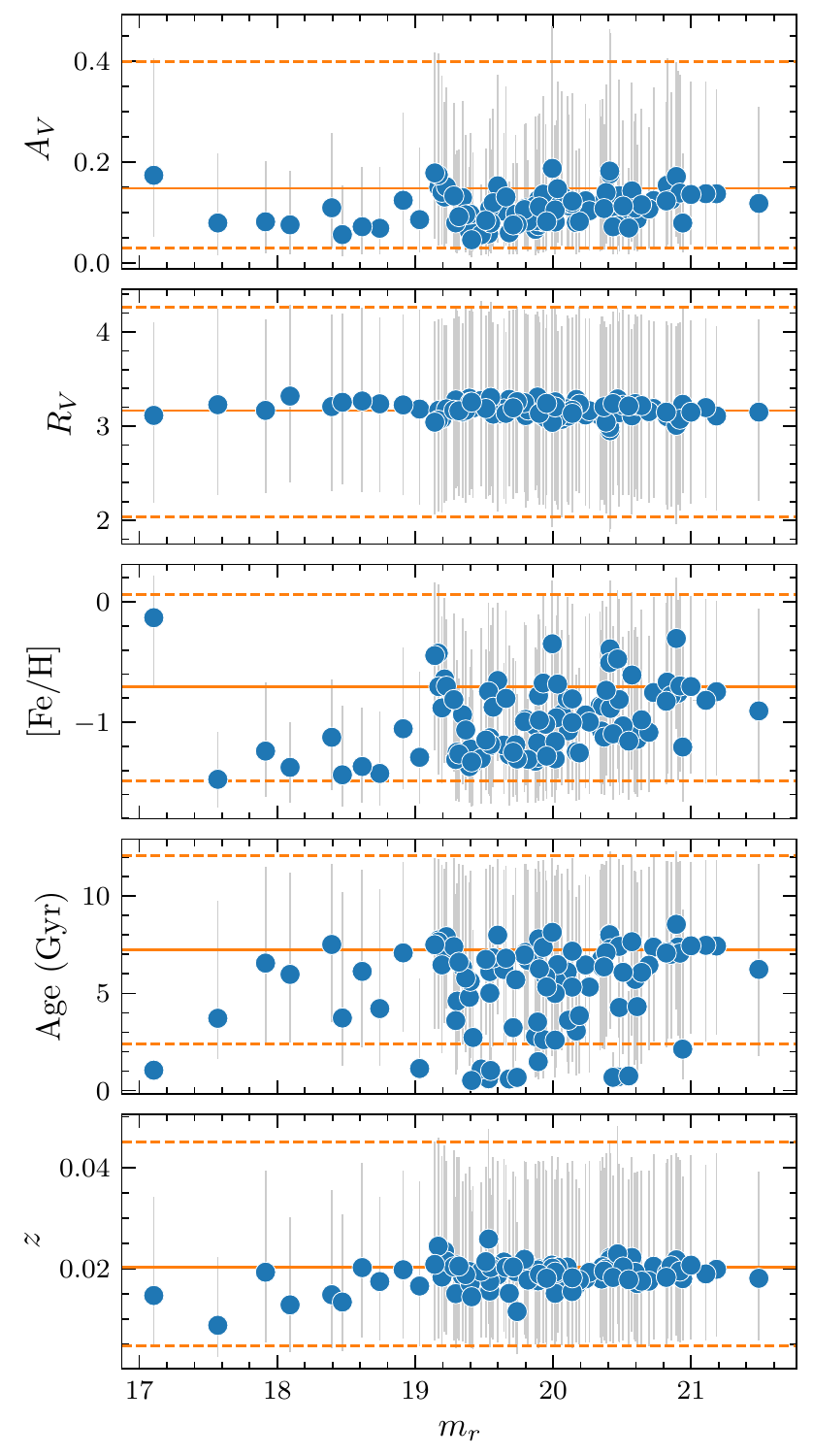}
\caption{SED fitting parameters as a function of the $r$-band apparent magnitude. The blue circles indicate the posterior distribution median and the grey vertical lines represents the $1\sigma$ uncertainties of the five free parameters, while the orange solid (dashed) lines indicate the median  ($\pm1\sigma$ uncertainties) of the prior distributions of the same parameters. \label{fig:magr_params}}
\end{figure}

To understand how well we constrain the parameters in our model, we compare the posterior distributions with the prior distributions. We perform this exercise in Fig.~\ref{fig:magr_params}, where we show the posterior medians and uncertainties for five free parameters in our model ($A_V$, $R_V$, Age, [Fe/H], $z$) as a function of $m_r$, the apparent magnitude of the galaxies according to the SMUDGes measurements. Overall, the fitting is better constrained, i.e., is restricted to a narrower range of values in the posterior distribution, for the brighter sources ($m_r\lesssim 19$), while the posterior distributions tend to be more similar to the prior distribution for the fainter sources ($m_r\gtrsim 19$). We discuss the results for the individual parameters below.

The extinction law parameters have limited impact on the optical SED shape of the UDGs, and no strong dust attenuation was required to fit the models. The median total extinction of $A_V\approx 0.1-0.2$ is recovered in all cases, whereas the total-to-selective extinction $R_V$ is mostly unchanged from the prior distribution. In practice, both parameters have the role of nuisance parameters in our analysis, as they are not of direct interest for this work, but are still taken into consideration in the analysis of the stellar populations parameters and the redshift.

The metallicity clearly departs from the prior distribution in most cases, with median metallicities systematically small ([Fe/H]$\approx -1$ dex). Even though the $1\sigma$ uncertainties remain similar to the prior for the faint UDGs, the posterior distributions for the metallicity are usually skewed towards low metallicities in most cases, and are not flat-shaped like the priors. The main concern in the derived metallicities occur for the more metal-poor galaxies, given that they are sometimes compatible with the lowest metallicity available in our SSP grid ([Fe/H]=$-$1.79). Without a larger grid of models, we can not rule out that some of these systems have even lower metallicities. Overall, however, we conclude that our metallicities estimates are well constrained by our SED fitting. 

Similarly, despite the large uncertainties for the faint UDGs, we do find that the luminosity-weighted ages tend to be smaller than the prior median (Age $\approx 7$ Gyr). One important point in the evaluation of the ages is that we can see more variation in the SED's of younger galaxies, in the sense that it is easier to differentiate between a 1 Gyr and a 2 Gyr old population than to differentiate between a 10 Gyr and a 12 Gyr old population. We see that effect in practice in our modeling in Fig.~\ref{fig:magr_params}, as the posterior distribution for galaxies with young ages are usually narrower than the prior distribution, while for those with old ages we tend to obtain relatively flat posteriors. Overall, we conclude that we are able to differentiate between young and old stellar populations in our UDG candidates, which is enough to allow a broad discussion of the formation of these systems.

Finally, the quality of the modeled photometric redshifts also depends on the apparent magnitude of the UDGs. The posterior redshift distribution for the faint UDGs is very similar to the prior distribution. In these cases, the quoted uncertainties in the redshift are around $0.8$ dex, which is simply the propagation of allowed prior range. On the other hand, the bright UDGs have a narrower range of redshifts in the posterior distribution, and their median redshifts are usually smaller than the prior median of $\approx 0.02$. However, even in these cases, the redshift estimate is very uncertain, and we are only able to constrain the photometric redshift with errors $\sigma_z \approx 0.01$ in the best cases. This has important implications in the classification of the UDG candidates, as we discuss below.

\subsection{Implications of the estimated redshifts to the classification of UDG candidates}

The most important implication of the redshift uncertainty is on the question of whether the UDG candidates are real UDGs, i.e., are they physical large, $R_e\geq 1.5$ kpc. We showed in \S\ref{sec:posteriors} that we are only able to minimally constrain photometric redshift for the bright UDG candidates ($m_r\lesssim 19$), and we rely on the prior distribution to estimate the redshift of the fainter UDG candidates. 

\begin{figure}[ht!]
\epsscale{1.2} % 2 columns
\plotone{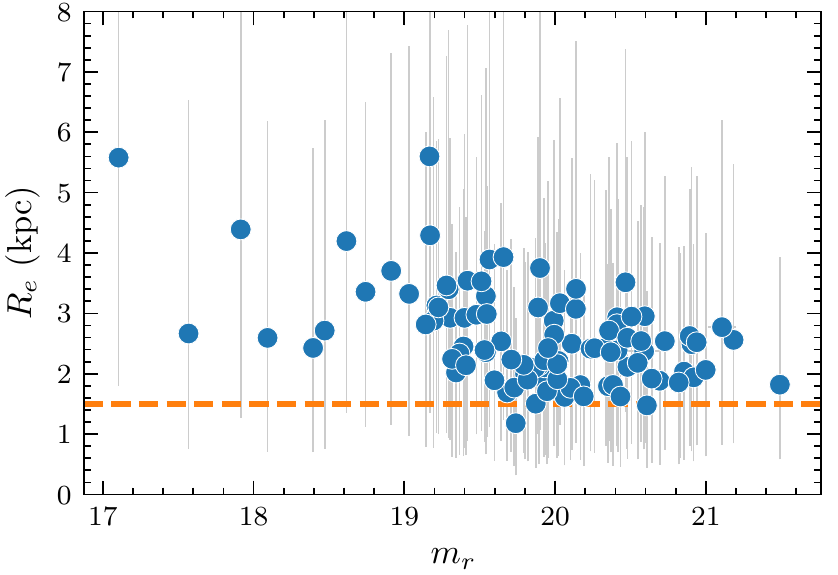}
\caption{Effective radii of UDG candidates as a function of their apparent magnitude. The circles and their uncertainties indicate the median and the $1\sigma$ uncertainties of the effective radius calculated using the posterior distributions of the SED fitting. The orange dashed line indicate the minimum physical radius of UDGs. \label{fig:magr_re}}
\end{figure}

If, for the sake of argument, we consider the photometric redshift estimates to be  correct, we can test whether the candidates can be classified as UDGs and whether this leads to any obvious irregularities. First, in Fig.~\ref{fig:magr_re} we show the estimated effective radii of our UDG candidates as a function of the apparent magnitude, using the posterior distribution samples for the photometric redshift, and adopting the angular sizes, $R_e$, determined by SMUDGes. We use the apparent magnitude as the independent variable to emphatize that our ability to constrain the sizes are directly affected by the observed luminosity of the UDGs, but this does not reflect the actual size-luminosity relation that is expected to exist for UDGs if they are similar to other galaxies \citep[e.g.][]{1977ApJ...218..333K}. All but two candidates are larger than the UDG criterion with greater than 50\% confidence. Of course, for the fainter systems this is principally a reflection of the adopted prior distribution, but for the brighter systems, where the determined redshift differs from the prior median, we have greater confidence that the physical sizes bear some resemblance to the truth. Second, the adopted redshifts do not lead to an unexpected set of very large ($R_e > 6$ kpc) UDGs. As such, our determinations are not manifestly incorrect.  

Given the limited redshift information contained in our observations, we are unable to conclude that our candidates are all real UDGs, but the bright ones are likely to be real UDGs, as well as the ones we discussed previously as confirmed through distance-by-association. For the sample as a whole, we argue based on volume considerations that they are likely to be farther away than our adopted median prior distance.
The argument goes as follows. First, we specify that the maximum size of any UDG is $R_e = 6$ kpc, which sets a maximum distance for each of our candidates. The candidate can lie at any distance up to this maximum distance. Second, we assume that the parent population of our candidates is uniformly distributed throughout the local volume. Third, we claim that our selection is effectively independent of distance, within this volume, because it depends on surface brightness more than on luminosity. The latter statement is not strictly correct, but valid at the coarse level of this argument \citep{2019ApJS..240....1Z}. In such a scenario the mean distance  to our candidates is $159 \pm 40$ Mpc, or $z = 0.036 \pm 0.01$, which is greater than our adopted median prior and supports the argument that the majority of candidates are indeed UDGs. In the next sections, we use the term UDG for all candidates, acknowledging that some of them might not meet the physical size criterion for UDGs. 

\section{Discussion}
\label{sec:discussion}

In this section we examine a variety of established galactic relations and properties, and place our UDG sample in context. We restrict our discussion to the stellar population properties and to only one variable that depends on the distance, the stellar mass, to avoid observed relations that may be contaminated by large correlations among parameters owing to our photometric redshift estimations.

\subsection{The stellar mass-metallicity relation of UDGs}
\label{sec:mass-metal}

We begin this exploration by determining whether \ac{UDGs} are similar to other \ac{LSB} galaxies \citep[see ][]{2012AJ....144....4M, 2013ApJ...779..102K}, and thus follow the same stellar mass-metallicity as bright galaxies \citep{2005MNRAS.362...41G}. Previous studies found that UDGs usually conform to the stellar mass-metallicity relation defined by dwarf galaxies, but most of those UDGs are in or near clusters, such as Coma \citep{2018ApJ...859...37G, 2018MNRAS.479.4891F, 2018MNRAS.478.2034R}, with a few examples of UDGs not associated to clusters \citep{2016AJ....151...96M, 2018ApJ...866..112G, 2019A&A...625A..77F}. 

\begin{figure}[ht!]
\epsscale{1.15} % 2 columns
\plotone{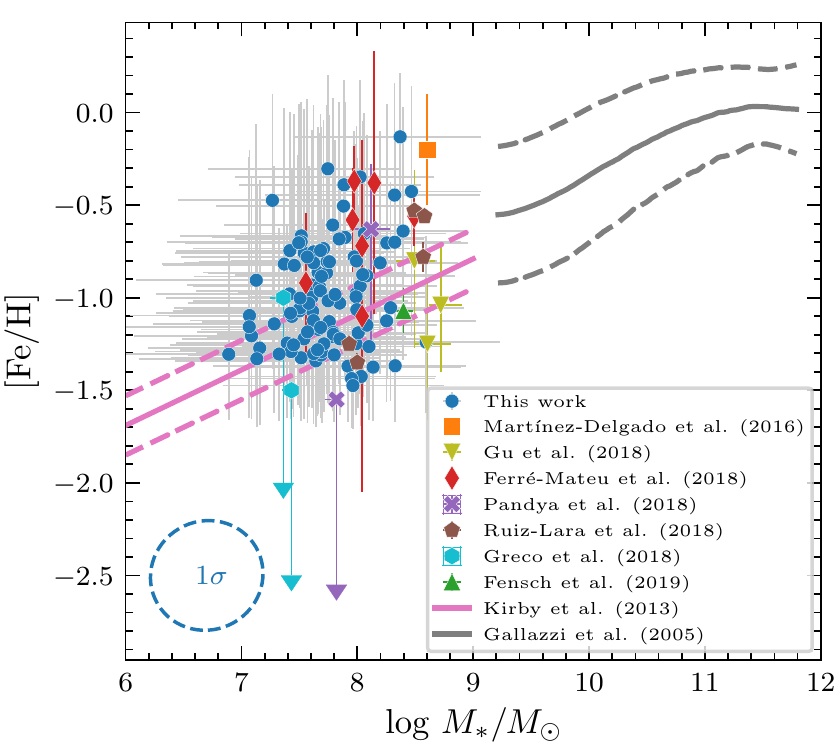}
\caption{The stellar mass-metallicity relation fThe blue dashed ellipse in the bottom right indicates the mean 1$\sigma$ covariance between the parameters, where the direction of largest (smallest) variance corresponds to the major (minor) semi-major axis.or UDGs in our sample (filled blue circles) and from the literature \citep{2016AJ....151...96M, 2018ApJ...859...37G, 2018MNRAS.479.4891F, 2018ApJ...858...29P, 2018MNRAS.478.2034R, 2018ApJ...866..112G, 2019A&A...625A..77F}. The pink solid and dashed lines are the mean and the scatter of the stellar mass-metallicity of dwarf galaxies around the Milky Way from \citet{2013ApJ...779..102K}. The gray lines are the mean and the standard deviation of the properties of bright galaxies from \citet{2005MNRAS.362...41G}. The shape of the blue dashed ellipse in the bottom left indicates the mean $1\sigma$ covariance between the parameters, where the direction of largest (smallest) variance corresponds to the major (minor) semi-major axis.\label{fig:stellarmass-metal}}
\end{figure}

In Figure \ref{fig:stellarmass-metal} we show the stellar mass-metallicity relation and include our sample of UDGs. There is a large variety of data types, models and methods involved in the determination of stellar populations of UDGs in the literature, and part of the scatter in the mass-metallicity and other relations may be attributed to that. For instance \citet{2018ApJ...866..112G} have shown a difference of $0.3-0.5$ dex in the metallicity of UDGs by simply changing from a \acl{SSP} to an extended star formation history. However, despite this important caveat, our measurements are consistent with previous work, with UDGs filling part of the gap between dwarf and giant galaxies. In the bottom left of the figure, we included an ellipse whose shape shows the covariance between the two parameters, which indicates that the stellar mass and the metallicity are basically independent in our measurements.

Our large sample of galaxies allows the observation of a range of metallicities that matches the variety of metallicities previously indentified in the literature. However, the large uncertainties and the censored limits in the range of metallicites do not allows a reliable  measure of the metallicity scatter for the UDGs in the sample. Overall, the location of the population of \ac{UDGs} in the stellar mass-metallicity diagram indicates a similarity with other dwarf \ac{LSB} galaxies, such as those observed by \citet{2013ApJ...779..102K}. On average, the metallicities of the UDGs, as presented, are slightly larger than those predicted from the extrapolation of the relation derived from measurements of dwarf galaxies, but there are a few important considerations that favor the idea that the metallicity of UDGs follows the same relation of the dwarf galaxies. 

First, the UDGs are not statistically far away from the dwarf sequence. Considering only our sample of UDGs, the mean difference between the measured metallicity and the expected metallicity from the  \citet{2013ApJ...779..102K} relation is 0.18 dex, which is similar to the scatter of the dwarf galaxies around the mean (0.14 dex), and much smaller than the mean error in our measurements (0.6 dex). Second, there may be a systematic error in our measurements related to the assumed relation between the total metallicity and the iron abundance, [M/H]=[Fe/H], because the low metallicity stars used in the E-MILES models contain alpha elements. \citet{2018MNRAS.479.4891F} reported a few UDGs with significant over abundance of magnesium ( $0\lesssim$ [Mg/Fe] $\lesssim0.4$ for three out of four galaxies) and \citet{2019MNRAS.484.3425M} reported an even larger over abundance in DGSAT I, [Mg/Fe]=1.5. 
An average magnesium abundance of [Mg/Fe]$\approx 0.2$ dex is  enough to account for the difference we find between the metallicity of our UDGs and that published for the dwarf galaxies\footnote{For the E-MILES models, [Fe/H]=[M/H]-0.75[Mg/Fe].}. Finally, our \ac{SSP} models are restricted to a lower limit of [Fe/H]$=-1.8$, and thus the metallicity of some UDGs in our sample may be over estimated. Note that a factor of two smaller distances, which would then render most of our candidates as non-UDGs, would lead to a factor of 4 lower stellar mass and would exacerbate the metallicity offset.

In conclusion, the metallicity of the UDGs is roughly consistent with that of other galaxies of similar stellar masses, the high-mass end of the dwarf sequence, and so do not show any signs of a different formation path than those galaxies. 

\subsection{The luminosity-weighted ages of field UDGs}

The reported ages of UDGs have usually been large, $>$ 4 Gyrs, but again almost all published results are for Coma galaxies \citep{2018ApJ...859...37G, 2018MNRAS.478.2034R, 2018MNRAS.479.4891F}. In the limited available examples of field UDGs, however, the reported luminosity-weighted ages have consistently been younger, with ages ranging from 1 to 3 Gyr \citep{2016AJ....151...96M, 2018ApJ...866..112G, 2019MNRAS.484.3425M}.  In fact, the UDG population in the field is expected to have a larger variety of colors than that of the clusters \citep{2017MNRAS.466L...1D} and there is observational support for this trend \citep{2016A&A...590A..20V, 2019MNRAS.488.2143P}. 

\begin{figure}[ht!]
\epsscale{1.15} % 2 columns
\plotone{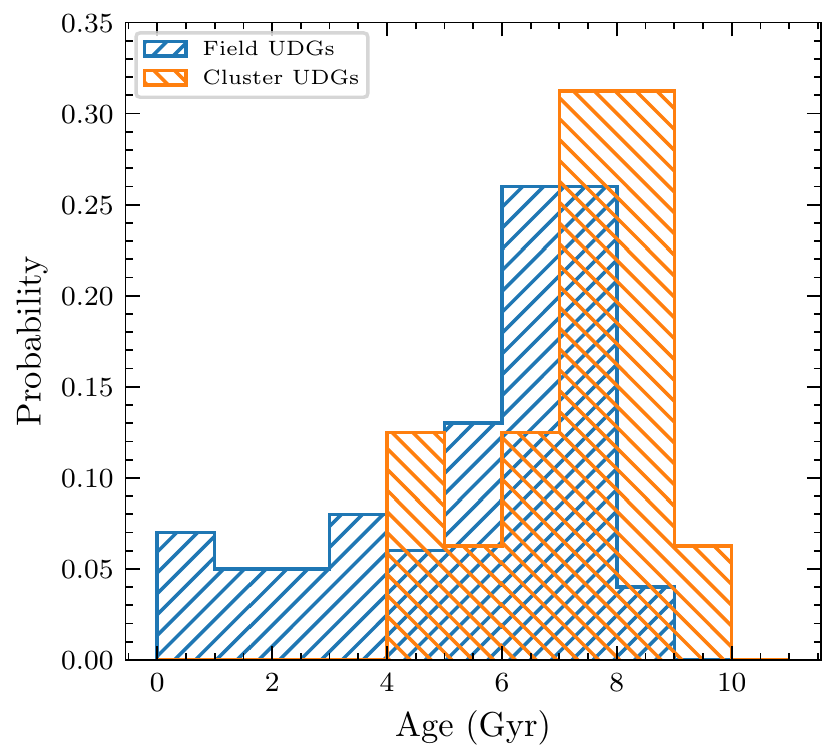}
\caption{Comparison of the luminosity-weighted ages of our \ac{UDGs}, which are primarily in the field, and those of UDGs in the literature \citep{2018ApJ...858...29P, 2018ApJ...859...37G, 2018MNRAS.478.2034R, 2018MNRAS.479.4891F}, which are primarily in clusters.\label{fig:age_hist}}
\end{figure}

In Figure \ref{fig:age_hist} we compare the luminosity-weighted age distribution of galaxies in our sample to that of UDGs observed in the Coma \citep{2018ApJ...859...37G,2018MNRAS.478.2034R,2018MNRAS.479.4891F} and Virgo \citep{2018ApJ...858...29P} clusters, to contrast the distribution of ages in clusters and in the field. Both the field and cluster UDGs typically have intermediate ages, with a peak in the age histogram around 7 Gyr, but our sample also indicates a significant fraction of UDGs with ages smaller than 4 Gyr.  

Considering that we only have luminosity-weighted ages, the results from the our analysis are expected to be biased toward the youngest populations within a galaxy. Therefore, a few different, non-exclusive scenarios can explain the additional fraction of UDGs with young ages. One possible explanation is the existence of different UDG formation channels not present in the cluster \citep[e.g.][]{2019MNRAS.tmp.2566L}. Other possibilities are that \ac{UDGs} in the field might have more continuous star formation activity, presumably in the absence of cluster-related processes, such as harassment and ram-pressure stripping, and that field UDGs have had a recent, even on-going, star formation burst that outshines the older and more massive stellar component of the galaxy. Regardless of the detailed explanation, UDGs are able to flourish in the field by forming stars until much more recently than UDGs in clusters. 

\subsection{The age-metallicity relation of UDGs}

\begin{figure}[ht!]
\epsscale{1.15} % 2 columns
\plotone{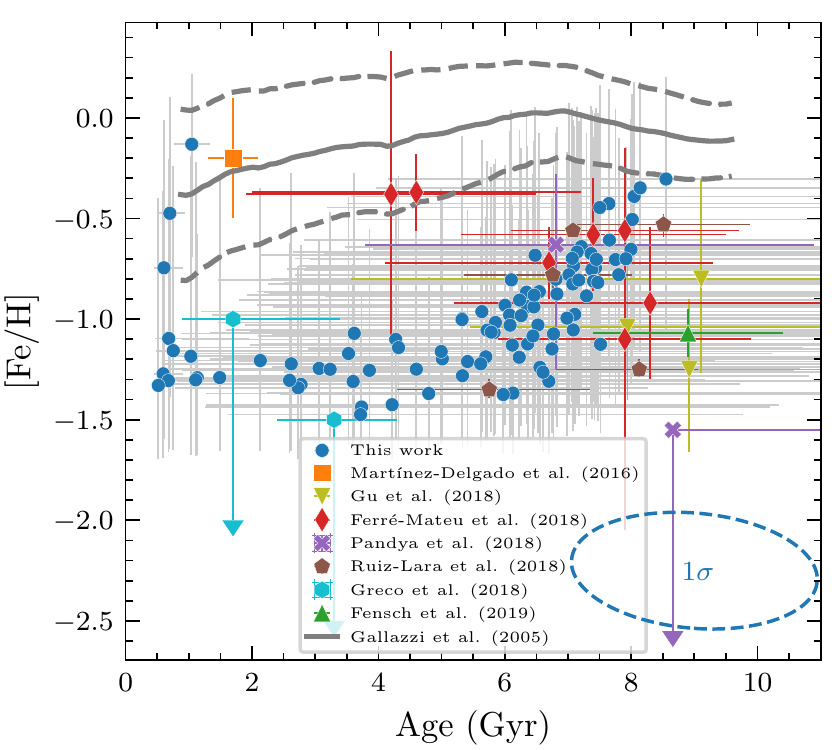}
\caption{Comparison between luminosity-weighted ages and metallicities of UDGs in the field (this work) and from the literature \citep{2016AJ....151...96M, 2018ApJ...859...37G, 2018MNRAS.479.4891F, 2018ApJ...858...29P, 2018MNRAS.478.2034R, 2018ApJ...866..112G, 2019A&A...625A..77F}. Solid and dashed lines are the mean and the standard deviation of the relation for bright galaxies \citep{2005MNRAS.362...41G}. The blue dashed ellipse in the bottom right indicates the mean 1$\sigma$ covariance between the parameters. \label{fig:age_metal}}
\end{figure}

We compare in Figure \ref{fig:age_metal} the luminosity-weighted age-metallicity relation for \ac{UDGs}, both in the field and in galaxy clusters, to that of bright galaxies. Within our own field UDG sample, there appears to be a correlation between age and metallicity, but considering the existence of a well-known age-degeneracy problem, we first inspect whether this is causing the observed relation.

The original age-deneracy problem \citep{1994ApJS...95..107W} indicates that the colors of an old population are similar to those of another population three times older and with half the metallicity. This degeneracy is specific for broad bands and old stellar systems, and thus it is unclear whether this holds in our analysis. However, it is an important cautionary note to any stellar population analysis, as degeneracies are bound to happen in SED fitting. As we indicate with the error ellipse in the bottom of the figure, there is a correlation between the age and the metallicity in our posterior distributions that is similar to the known age-metallicity degeneracy. However, the observed relation between the ages and metallicities of our UDGs does not have a slope in the same direction as the age-degeneracy relation. Therefore, we conclude that the observed relation between ages and metallicities in our UDG sample is not driven by the age-metallicity degeneracy, and thus we are able to discuss some properties of the observed relation.

For the old UDGs (age$\gtrsim 6$ Gyr), the age-metallicity relation follows a similar pattern to that of bright galaxies \citep{2005MNRAS.362...41G}, in the sense that younger systems have low metallicity and older systems have high metallicity, although with different slope and offset. The old UDGs have metallicities similar to those reported in other works for UDGs in clusters, but this possible age-metallicity relation was not hinted at in previous work.

The young UDGs have a flat age-metalicity relation, but the modeling limitation in the range of very low metallicities limits us in reaching further conclusions as to whether the flattening in the relation is real or a result of the modeling restriction. The location of our young UDGs in this space is similar to that of the field UDGs from 
\citet{2018ApJ...866..112G}, which were suggested to be currently star forming.  

Interestingly, there are also a few young UDGs ($t\lesssim 1$ Gyr) with relatively high metallicities ([Fe/H]$\approx -0.5$ dex), populating the location of more massive galaxies. These UDGs are located in age-metallicity plane in a location similar to that of DGSAT I, a passive, field UDG found in the  Pisces-Perseus supercluster filament \citep{2016AJ....151...96M}. A visual inspection of our young, metal-rich UDGs does not suggest current tidal interactions with bright galaxies, and thus it is not likely that these particular UDGs have tidal origins, which could have explained their higher metallicity. A more likely scenario is that these cases indicate more massive UDGs that have a recent burst in star formation, such that their luminosity-weighted metallicities are driven by a potentially old, mass-dominant stellar component, while their luminosity-weighted ages are driven by a less massive, young component.

\subsection{The stellar mass-age relation}

\begin{figure}[ht!]
\epsscale{1.15} % 2 columns
\plotone{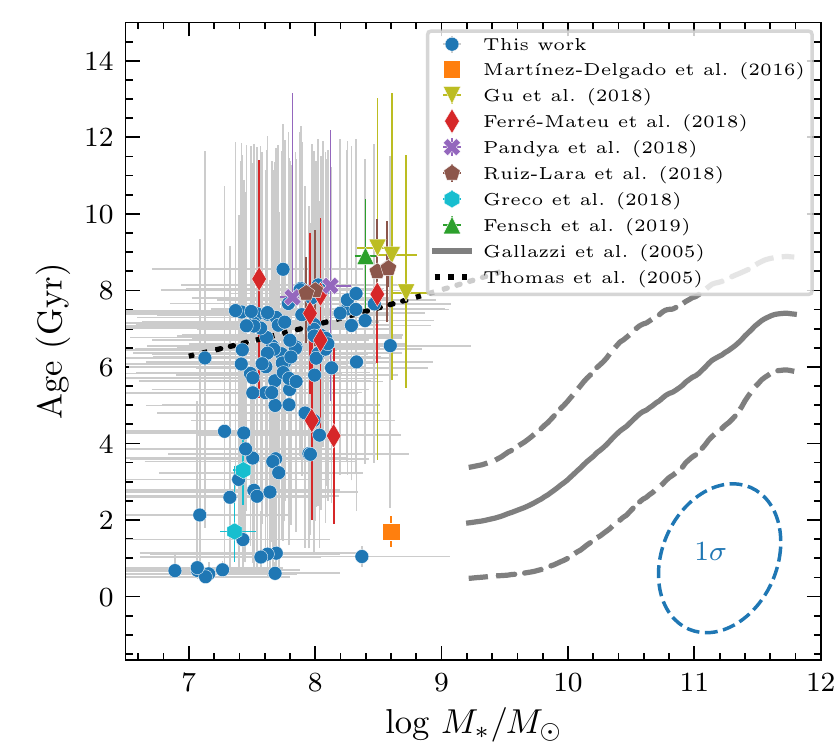}
\caption{Same as Fig.~\ref{fig:age_metal} for the relation between stellar mass and luminosity-weighted age for UDGs and bright galaxies. \label{fig:mass_age}}
\end{figure}

In Figure~\ref{fig:mass_age} we present the relation between the stellar mass and the luminosity-weighted age for our UDG sample and more massive galaxies. Similarly to the age-metallicity relation, we also observe a correlation between the ages and the stellar masses. However, in this case, the error ellipse in the bottom of the figure indicates that the observed correlation may be caused by a degeneracy in the parameters, and thus we do not have any confidence that this relation actually exists.

Most old UDGs (Age $\gtrsim 6$ Gyr) follow the stellar mass - mass weighted age relation observed by \citet{2005ApJ...621..673T}, extrapolated to the UDG regime. We explain this agreement by noting that the luminosity-weighted and mass-weighted ages converge the longer a galaxy is not forming stars. However, the young UDGs are displaced from the relation of \citet{2005ApJ...621..673T}, and have ages similar to those in the low-mass end of the \citet{2005MNRAS.362...41G} relation. We expect these young field UDGs to move upward in this diagram when they eventually stop forming stars.

\section{Summary and conclusion}
\label{sec:conclusion}

Ultra-diffuse galaxies (UDGs) are large, low surface brightness galaxies. Although such systems are now known in significant numbers, establishing physical characteristics has proven to be challenging even when using the largest telescopes of this generation. Field UDGs, in particular, have barely been studied. In this work, we perform the first systematic study of the stellar populations of field \ac{UDGs} combining the deep and large area search of UDGs performed by the SMUDGes survey \citep{2019ApJS..240....1Z} with the multiband capabilities of the  \ac{S-PLUS} survey \citep{2019MNRAS.489..241M}. Covering an area of the $\sim$330 deg$^2$ in the Stripe 82 region, we fit spectral energy distributions (SEDs) to a sample of 100 field UDGs, representing the largest sample of UDGs (field or cluster) for which ages and metallicites have been measured to date. 

We constrain our Bayesian SED fitting method with up to 12 broad and narrow bands from S-PLUS, resulting in estimated luminosity-weighted ages, metallicities and stellar masses of the field UDGs. While stellar masses and metallicities are mostly in agreement with previous studies, we observed a number of UDGs with ages younger than those found in cluster. This result suggests that UDGs in the field may have extended star formation histories that may, in some cases, extend to the current time, contrasting with the typical old ages of UDGs found in clusters. We also found a few cases of relatively high-metallicity UDGs ([Fe/H]$\approx -0.5$) with young ages (ages $\lesssim 1$ Gyr) without clear indications of tidal interactions, which suggest ongoing episodes of star formation among the most massive UDGs.

Previous studies have already indicated that UDGs may represent the extension of normal galaxy properties rather than arising from a disconnected, new path of galaxy formation, but these conclusions have been based on small samples of galaxies \citep{2016ApJ...830...23B, 2017MNRAS.464L.110Z} or models \citep{2016MNRAS.459L..51A}. By placing a large sample of field UDGs in stellar population scaling relations, we were able to confirm that UDGs, as a population, are similar to dwarf and giant galaxies. There are still puzzles to solve, such as the large globular cluster abundances in the largest UDGs \citep{2017ApJ...844L..11V, 2018ApJ...856L..31T} and the offset from the baryonic Tully-Fisher relation \citep{2019ApJ...883L..33M}, but we conclude that these should arise naturally from considering a broader range of galaxies within the current picture of galaxy formation \citep{2019MNRAS.485..796M} rather than any exotic processes \citep{2018ApJ...866L..11B}. Of course, these statements apply to the general case and individual exceptions, where UDGs form in tidal tails, for example, are not excluded.

Despite the improvement in sample size in this work, there is still much to be gained from even larger samples. In particular, we want to apply the same analysis methods to UDGs in a range of environments, including massive clusters, to more confidently compare results. Even larger samples will enable us to test for further dependencies on UDG mass, environment, and morphology. Both SMUDGes and S-PLUS are still in their early phases. A much larger area of the sky will be probed by both surveys in the next years, increasing the sample for which this type of analysis can be replicated into the thousands.

\bibliography{smudges-splus}{}
\bibliographystyle{aasjournal}

\acknowledgments

The authors thank the anonymous referee for his/her comments. We are thankful to Stavros Akras, Yoli Jim\'enez Teja, Marco Grossi, Alvaro Alvarez-Candal, Jos\'e Luis Nilo Castell\'on, Paulo Lopes, Kanak Saha, Eduardo Telles and Ana Chies Santos for providing comments and suggestions. CEB, CMdO gratefully acknowledges the S\~ao Paulo Research Foundation (FAPESP), grants 2011/51680-6, 2016/12331-0 and 2018/24389-8. DZ, RD, and HZ gratefully acknowledge financial support from NSF AST-1713841. PC acknowledges support from FAPESP project 2018/05392-8, and Conselho Nacional de Desenvolvimento Cient\'ifico e Tecnol\'{o}gico (CNPq) project 310041/2018-0. LS thanks the FAPESP scholarship grant 2016/21664-2. FRH thanks FAPESP for the financial support, grants 2019/23141-5 and 2018/21661-9. J.\,A.\,H.\,J. thanks to Brazilian  institution CNPq for financial support through  postdoctoral fellowship (project 150237/2017-0) and Chilean institution CONICYT, Programa de Astronom\'ia, Fondo ALMA-CONICYT 2017, C\'odigo de proyecto 31170038. The T80­South robotic telescope \citep{2019MNRAS.489..241M} was founded as a partnership between FAPESP, the Observat\'orio Nacional (ON), the Federal University of Sergipe (UFS) and the Federal University of Santa Catarina (UFSC), with important financial and practical contributions from other collaborating institutes in Brazil, Chile (Universidad de La Serena) and Spain (CEFCA). This work has made use of the computing facilities of the Laboratory of Astroinformatics (Instituto de Astronomia, Geof\'isica e Ci\^encias Atmosf\'ericas, Departamento de Astronomia/USP, NAT/Unicsul), whose purchase was made possible by FAPESP (grant 2009/54006-4) and the INCT-A. This research has made use of the NASA/IPAC Extragalactic Database (NED), which is operated by the Jet Propulsion Laboratory, California Institute of Technology, under contract with the National Aeronautics and Space Administration.


\vspace{5mm}
\facilities{T80South}


\software{astropy \citep{2013A&A...558A..33A},  
          matplotlib \citep{4160265},
          numpy \citep{5725236},
          pymc3 \citep{Salvatier2016},
          scipy \citep{scipy},
          % SExtractor \citep{1996A&AS..117..393B},
          photutils \citep{Bradley_2019_2533376}.}



\end{document}